\documentclass[12pt]{article}
\usepackage{amsmath,amssymb,amscd,mathtools}
\usepackage{algorithm}
\usepackage{algpseudocode}
\newlength{\maxwidth}
\newcommand{\algalign}[2]{\makebox[\maxwidth][r]{#1{}}{}#2}

\usepackage{times}
\usepackage{graphicx}
\usepackage{color}
\usepackage{multirow}
\usepackage[authoryear]{natbib}
\usepackage{rotating}
\usepackage{bbm}
\usepackage{latexsym}
\usepackage{enumerate}
\usepackage{hyphenat}
\usepackage[dvipsnames]{xcolor}

\newcommand{\captionfonts}{\normalsize}

\newcommand{\DKL}{\mathrm{D_{KL}}}
\newcommand{\h}{\hspace{1pt}}

\makeatletter  
\long\def\@makecaption#1#2{%
  \vskip\abovecaptionskip
  \sbox\@tempboxa{{\captionfonts #1: #2}}%
  \ifdim \wd\@tempboxa >\hsize
    {\captionfonts #1: #2\par}
  \else
    \hbox to\hsize{\hfil\box\@tempboxa\hfil}%
  \fi
  \vskip\belowcaptionskip}
\makeatother   
\usepackage{caption}
\begin{document}

\date{}
\title{\bf Systems of bounded rational agents with information\hyp theoretic constraints}
\maketitle

{\bf \large Sebastian Gottwald$^{\displaystyle 1}$ and \bf \large Daniel A. Braun$^{\displaystyle 1}$}\\
{$^{\displaystyle 1}$Institute of Neural Information Processing, Faculty of Engineering, Computer Science and Psychology, University of Ulm.}\\
%

%\ \\[-2mm]
{\bf Keywords: } Bounded rationality, multi-agent systems, hierarchical structure, specialization, Free Energy principle 

\thispagestyle{empty}
\markboth{S. Gottwald and D. A. Braun}{Systems of bounded rational agents with information\hyp theoretic constraints}
\ \vspace{-0mm}\\
%
%Abstract
\begin{abstract}
Specialization and hierarchical organization are important features of efficient collaboration in economical, artificial, and biological systems. Here, we investigate the hypothesis that both features can be explained by the fact that each entity of such a system is limited in a certain way. We propose an information-theoretic approach based on a Free Energy principle, in order to computationally analyze systems of bounded rational agents that deal with such limitations optimally. We find that specialization allows to focus on fewer tasks, thus leading to a more efficient execution, but in turn requires coordination in hierarchical structures of specialized experts and coordinating units. Our results suggest that hierarchical architectures of specialized units at lower levels that are coordinated by units at higher levels are optimal, given that each unit's information-processing capability is limited and conforms to constraints on complexity costs. %%%%%%%%%%%

\end{abstract}

\section{Introduction} \label{sec:intro}
The question of how to combine a given set of individual entities in order to perform a certain task efficiently is a long-lasting question shared by many disciplines, including economics, neuroscience, and computer science. Even though the explicit nature of a single individuum might differ between these fields, e.g. an employee of a company, a neuron in a human brain, or a computer or processor as part of a cluster, they have one important feature in common that usually prevents them from functioning isolated by themselves: they are all limited. In fact, this was the driving idea that inspired Herbert A. Simons early work on decision-making within economic organizations \citep{Simon1943,Simon1955}, which earned him a Nobel prize in 1978. He suggested that a scientific behavioral grounding of economics should be based on bounded rationality, which has remained an active research topic until today \citep{Russel1995,Lipman1995,Aumann1997,Kaelbling1998,DeCanio1998,Gigerenzer2001,Jones2003,Sims2003,Burns2013,Ortega2013,Acerbi2014,Gershman2015}. 
Subsequent studies in management theory have been built upon Simons basic observation, because ``if individual managers had unlimited access to information that they could process costlessly and instantaneously, there would be no role for organizations employing multiple managers" \citep{Geanakoplos1991}. In neuroscience and biology, similar concepts have been used to explore the evolution of specialization and modularity in nature \citep{Kashtan2005,Wagner2007}. In modern computer science, the terms parallel computing and distributed computing denote two separate fields that share the concept of decentralized computing \citep{Radner1993}, i.e. the combination of multiple processing units in order to decrease the time of computationally expensive calculations.

Despite of their success, there are also shortcomings of most approaches to the organization of decision-making units based on bounded rationality: As \citep{DeCanio1998} point out, existing agent-based methods (including their own) are not using an overreaching optimization principle, but are tailored to the specific types of calculations the agents are capable of, and therefore lack in generality. Moreover, it is usually imposed as a separate assumption that there are two types of units, specialized operational units and coordinating non-operational units, which was expressed by \citep{Knight1921} as ``workers do, and managers figure out what to do''. 

Here, we use a Free Energy optimization principle in order to study systems of bounded rational agents, extending the work in \citep{Ortega2011,Ortega2013,Genewein2013,Genewein2015} on decision-making, hierarchical information\hyp processing, and abstraction in intelligent systems with limited information-processing capacity, that has precursors in the economic and game-theoretic literature \citep{McKelvey1995,Ochs1995,Mattsson2002,Wolpert2006,Spiegler2011,Howes2009,Todorov2009,Still2009,Tishby2011,Kappen2012,Edward2014,Lewis2014}. {\color{Black} Note that the Free Energy optimization principle of information-theoretic bounded rationality is connected to the Free Energy principle used in variational Bayes and Active Inference \citep{Friston20152,Friston2015,Friston20172,Friston2017}, but has a conceptually distinct interpretation and some formal differences (see Section \ref{bayesmodelsel} for a detailed comparison).}

By generalizing the ideas in \citep{Genewein2013,Genewein2015} on two-step information-processing to an arbitrary number of steps, we arrive at a general Free Energy principle that can be used to study systems of bounded rational agents. The advantages of our approach can be summarized as follows:

\begin{enumerate}[$(i)$]
\item There is a unifying Free Energy principle that allows for a multi-scale problem formulation for an arbitrary amount of agents distributed among the steps of general multi-step processes (see Sections \ref{sec:nodebasedFE} and \ref{sec:FE_ag}).

\item The computational nature of the optimization principle allows to explicitly calculate and compare optimal performances of different agent architectures for a given set of objectives and resource constraints (see Section \ref{sec:optimalarchs}).

\item The information-theoretic description implies the existance of the two types of units mentioned above, non-operational units (selector nodes) that coordinate the activities of operational units. Depending on their individual resource constraints, the Free Energy principle assigns each unit to a region of specialization that is part of an optimal partitioning of the underlying decision space (see Section \ref{sec:spec}).
\end{enumerate}

In particular, we find that, for a wide range of objectives and resource limitations (see Sections \ref{sec:objectives} and \ref{sec:bounds}), hierarchical systems with specialized experts at lower levels and coordinating units at higher levels generally outperform other structures.

\bigskip

\section{Preliminaries} \label{sec:prelim}

This section serves as an introduction to the terminology required for our framework presented in Section 3 and 4.

{\color{Black}
\subsection*{Notation} 
We use curly letters, $\mathcal W$, $\mathcal X$, $\mathcal A$, etc.~to denote sets of finite cardinality, in particular the underlying spaces of the corresponding random variables $W$, $A$, $X$, etc., whereas the values of these random variables are denoted by small letters, i.e. $w\in\mathcal W$, $a\in\mathcal A$, and $x\in\mathcal X$, respectively. We denote the space of probability distributions on a given set $\mathcal X$ by $\mathbb P_{\mathcal X}$. 
Given a probability distribution $p\in\mathbb P_{\mathcal X}$, the expectation of a function $f:\mathcal X \to\mathbb R$ is denoted by $\langle f\rangle_p \coloneqq \sum_x p(x) f(x)$. If the underlying probability measure is clear without ambiguity we just write $\langle f \rangle$.

For a function $g$ with multiple arguments, e.g. for $g:\mathcal X\times \mathcal Y \,{\to}\, \mathbb R, (x,y) \,{\mapsto}\, g(x,y)$, we denote the function $\mathcal X \,{\to}\, \mathbb R, x\,{\mapsto}\, g(x,y)$ for fixed $y\in\mathcal Y$ by $g(\h\cdot\h,y)$ (partial application), i.e.~the dot indicates the variable of the new function. Similarly, for fixed $y\in\mathcal Y$, we denote a conditional probability distribution on $\mathcal X$ with values $p(x|y)$ by $p(\h \cdot \h|y)$. This notation shows the dependencies clearly without giving up the original function names and thus allows to write more complicated expressions in a concise form. For example, if $F$ is a functional defined on functions of one variable, e.g. $F[f] \coloneqq \sum_{x} f(x)$ for all functions $f:\mathcal X\to \mathbb R$, then evaluating $F$ on the function $g$ in its first variable while keeping the second variable fixed, is simply denoted by $F[g(\cdot,y)]$. Here, the dot indicates on which argument of $g$ the functional $F$ is acting and at the same time it records that the resulting value (which equals $\sum_{x} g(x,y)$ in the case of the example) does not depend on a particular $x$ but on the fixed $y$.

}

\subsection{Decision-making} Here, we consider (multi-task) \textit{decision-making} as the process of observing a \textit{world state} $w\in\mathcal W$, sampled from a given distribution $\rho\in\mathbb P_{\mathcal W}$, and choosing a corresponding \textit{action} $a\in\mathcal A$ drawn from a \textit{posterior} policy $P(\h\cdot\h|w)\in \mathbb P_{\mathcal A}$. Assuming that the joint distribution of $W$ and $A$ is given by $p(a,w)\coloneqq \rho(w)P(a|w)$, then $P$ is the conditional probability distribution of $A$ given $W$. Unless stated otherwise, the capital letter $P$ always denotes a posterior, while the small letter $p$ denotes the joint distribution or a marginal of the joint (i.e. a dependent variable).

A decision-making unit is called \textit{agent}. An agent is \textit{rational}, if its posterior policy $P$ maximizes the \textit{expected utility}
\begin{equation}
\langle U\rangle = \sum_{w\in\mathcal W} \rho(w) \sum_{a\in\mathcal A} P(a|w)\,U(a,w) \label{EU}
\end{equation}
for a given \textit{utility function} $U:\mathcal W\times\mathcal A \to \mathbb R$. {\color{Black}Note that the utility $U$ may itself represent an expected utility over consequences in the sense of  \cite{Neumann1944}, where $W$ would serve as a context variable for different tasks.
The posterior $P$ can be seen as a state-action policy that selects the best action $a\in\mathcal A$ with respect to a utility function $U$ given the state $w\in\mathcal W$ of the world.}

\subsection{Bounded rational agents}\label{sec:bragents} In the information-theoretic model of bounded rationality \citep{Ortega2011,Ortega2013,Genewein2015}, an agent is \textit{bounded rational} if its posterior $P$ maximizes \eqref{EU} subject to the constraint
\begin{equation}
\big\langle \DKL(P\|q)\big\rangle \, = \, \sum_{w\in\mathcal W} \rho(w)\, \DKL(P(\h\cdot\h|w)\|q) \, \leqslant \, D_0 \, , \label{DKL}
\end{equation}
for a given bound $D_0>0$ and a \textit{prior} policy $q\in \mathbb P_{\mathcal A}$. Here, $\DKL(p\|q)$ denotes the Kullback-Leibler (KL) divergence between two distributions $p,q\in\mathbb P_{\mathcal Y}$ on a set $\mathcal Y$, defined by $\DKL(p\|q) \coloneqq \sum_{y\in \mathcal Y} p(y) \log (p(y)/q(y))$. Note that, for $\DKL(p\|q)$ to be well-defined, $p$ must be absolutely continuous with respect to $q$, so that $q(y)=0$ implies $p(y)=0$. When $p$ or $q$ are conditional probabilities, then we treat $\DKL(p\|q)$ as a function of the additional variables.

Given a world state $w$, the information-processing consists of transforming a prior $q$ to a world state specific posterior distribution $P(\h\cdot\h|w)$. Since $\DKL(P(\h\cdot\h|w)\|q)$ measures by how much $P(\h\cdot\h|w)$ diverges from $q$, the upper bound $D_0$ in \eqref{DKL} characterizes the limitation of the agent's average \textit{information-processing capability}: If $D_0$ is close to zero, the posterior must be close to the prior for all world states, which means that $A$ contains only little information about $W$, whereas if $D_0$ is large, the posterior is allowed to deviate from the prior by larger amounts and therefore $A$ contains more information about $W$. We use the KL-divergence as a proxy for any resource measure, as any resource must be monotone in processed information, which is measured by the KL-divergence between prior and posterior.

{\color{Black}
Technically, maximizing expected utility under the constraint \eqref{DKL} is the same as minimizing expected complexity cost under the constraint of a minimal expected performance, where complexity is given by the expected KL-divergence between prior and posterior and performance by expected utility. Minimizing complexity means minimizing the number of bits required to generate the actions.
}

\subsection{Free Energy principle} By the variational method of Lagrange multipliers, the above constrained optimization problem is equivalent to the unconstrained problem
\begin{equation} \label{FE}
\max_{P} \Big( \langle U\rangle - \frac{1}{\beta} \, \big\langle\DKL(P\|q)\big\rangle \Big) \ ,
\end{equation}
where $\beta>0$ is chosen such that the constraint \eqref{DKL} is satisfied. {\color{Black}In the literature on information-theoretic bounded rationality \citep{Ortega2011, Ortega2013}, the objective in \eqref{FE} is known as the \textit{Free Energy} $\mathcal F$ of the corresponding decision-making process.} In this form, the optimal posterior can be explicitly derived by determining the zeros of the functional derivative of $\mathcal F$ with respect to $P$, yielding the Boltzmann-Gibbs distribution
\begin{equation} \label{boltzmann}
P(a|w) = \frac{1}{Z(w)} \, q(a) \, e^{\beta \, U(a,w)} \, , \quad Z(w) \coloneqq \sum_{a\in\mathcal A} q(a) \, e^{\beta \, U(a,w)} \, .
\end{equation}
Note how the Lagrange multiplier $\beta$ (also known as \textit{inverse temperature}) interpolates between an agent with zero processing capability that always acts according to its prior policy ($\beta=0$) and a perfectly rational agent ($\beta\to\infty$). Note that, plugging \eqref{boltzmann} back into the Free Energy \eqref{FE} gives
\begin{equation} \label{optFE}
\max_P \mathcal F[P] = \frac{1}{\beta} \big \langle \log Z \big \rangle \, .
\end{equation}

\subsection{Optimal prior} \label{subsec:optprior}The performance of a given bounded rational agent crucially depends on the choice of the prior policy $q$. Depending on $D_0$ and the explicit form of the utility function, it can be advantageous to a priori prefer certain actions over others. Therefore, \textit{optimal} bounded rational decision-making includes optimizing the prior in \eqref{FE}. In contrast to \eqref{FE}, the modified optimization problem 
\begin{equation} \label{FE:opr}
\max_{P,q} \Big( \langle U\rangle - \frac{1}{\beta} \, \big\langle \DKL(P\|q)\big\rangle \Big) \ 
% \max_{P,q} \Big(\mathbb E_P[U] - \frac{1}{\beta} \mathbb E_\rho\big[\DKL(P\|q)\big] \Big) \, .
\end{equation}
does not have a closed form solution. However, since the objective is convex in $(P,q)$, a unique solution can be obtained iteratively by alternating between fixing one and optimizing the other variable \citep{Csiszar1984}, resulting in a Blahut-Arimoto type algorithm \citep{Arimoto1972,Blahut1972} that consists of alternating the equations
\begin{equation} \label{RD}
\left\{ \begin{array}{rl}P(a|w) \hspace{-6pt} & = \,  \frac{1}{Z(w)} \, q(a)\, e^{\beta U(a,w)}  \, ,\\[5pt]
q(a) \hspace{-6pt} & =\, p(a)  = \, \sum_w \rho(w) P(a|w), \end{array} \right. 
\end{equation}
with $Z(w)$ given by \eqref{boltzmann}. In particular, the optimal prior policy is the marginal $p$ of the joint distribution of $W$ and $A$. In this case, the average Kullback-Leibler divergence between prior and posterior coincides with the \textit{mutual information} between $W$ and $A$, 
\[
I(W;A) = \sum_{w\in\mathcal W} \sum_{a\in\mathcal A} p(w,a) \, \log \frac{p(w,a)}{\rho(w) p(a)} = \big\langle \DKL (P,p)\big\rangle \ .
\]
It follows that the modified optimization principle \eqref{FE:opr} is equivalent to 
\begin{equation}\label{FE:RD}
\max_{P} \Big( \langle U\rangle - \frac{1}{\beta}\, I(W;A) \Big) \, .
\end{equation}

Due to its equivalence to rate distortion theory \citep{Shannon1959} (with a negative distortion measure given by the utility function), \eqref{FE:RD} is denoted as the rate distortion case of bounded rationality in \citep{Genewein2013}.

\subsection{Multi-step and multi-agent systems} \label{sec:singletomulti}

When multiple random variables are involved in a decision-making process, such a process constitutes a \textit{multi-step system} (see Section \ref{sec:multistep}).
Consider the case of a prior over $\mathcal A$ that is conditioned on an additional random variable $X$ with values $x\in\mathcal X$, i.e. $q(\h\cdot\h|x)\in \mathbb P_{\mathcal A}$ for all $x\in\mathcal X$. Remember that we introduced a bounded rational agent as a decision-making unit, that, after observing a world state $w$,  transforms a single prior policy over a choice space $\mathcal A$ to a posterior policy $P(\h\cdot\h|w)\in\mathbb P_{\mathcal A}$. Therefore, in the case of a conditional prior, the collection of prior policies $\{q(\h\cdot\h|x)\}_{x\in\mathcal X}$ can be considered as a \textit{collection} {\color{Black}or \textit{ensemble of agents}}, or a \textit{multi-agent system}, where for a given $x\in \mathcal X$, the prior $q(\h\cdot\h|x)$ is transformed to a posterior $P(\h\cdot\h|x,w)\in\mathbb P_{\mathcal A}$ by exactly one agent. Note that a single agent deciding about both, $X$ and $A$, would be modelled by a prior of the form $q(x,a)$ with $x\in\mathcal X$ and $a\in\mathcal A$, instead.

Hence, in order to combine multiple bounded rational agents, we are first splitting the full decision-making process into multiple steps by introducing additional intermediate random variables (Section \ref{sec:multistep}), which then will be used to assign one or more agents to each of these steps (Section \ref{sec:multi-agent}). {\color{Black}In this view, we can regard a multi-agent decision-making system as performing a sequence of successive decision steps until an ultimate action is selected.}

\bigskip

\section{Multi-step bounded rational decision-making} \label{sec:multistep}

\subsection{Decision nodes} Let $W$ and $A$ denote the random variables describing the full decision-making process for a given utility function $U:\mathcal W\times\mathcal A\to\mathbb R$, as described in Section \ref{sec:prelim}. In order to separate the full process into $N>1$ steps, we introduce internal random variables $X_1$, \dots, $X_{N-1}$, which represent the outputs of additional intermediate bounded rational decision-making steps. For each $k$, let $\mathcal X_k$ denote the target space and $x_k\in\mathcal X_k$ a particular value of $X_k$. We call a random variable that is part of a multi-step decision-making system a \textit{(decision) node}. {\color{Black}For simplicity, we assume that all intermediate random variables are discrete (just like $W$ and $A$).}

Here, we are treating feed-forward architectures originating at $X_0\coloneqq W$ and terminating in $X_N\coloneqq A$. This allows to label the variables $\{X_k\}_{k=0}^{N}$ according to the information flow, so that $X_j$ potentially can only obtain information about $X_i$ if $i<j$. The canonical factorization 
\[
p(w,x_1,\dots,x_{N-1},a) = \rho(w)\, p(x_1|w) \, p(x_2|x_1,w) \cdots p(a|x_{N-1},\dots,x_1,w) \,
\]
of the joint probability distribution of $\{X_k\}_{k=0}^{N}$ therefore consists of the posterior policies of each decision node.

\subsection{Two types of nodes: inputs and prior selectors}
\begin{figure}
%\nopagenumber
%\renewcommand{\baselinestretch}{1.0}
\hfill
\noindent\makebox[\textwidth]{\includegraphics[width = 0.8\textwidth]{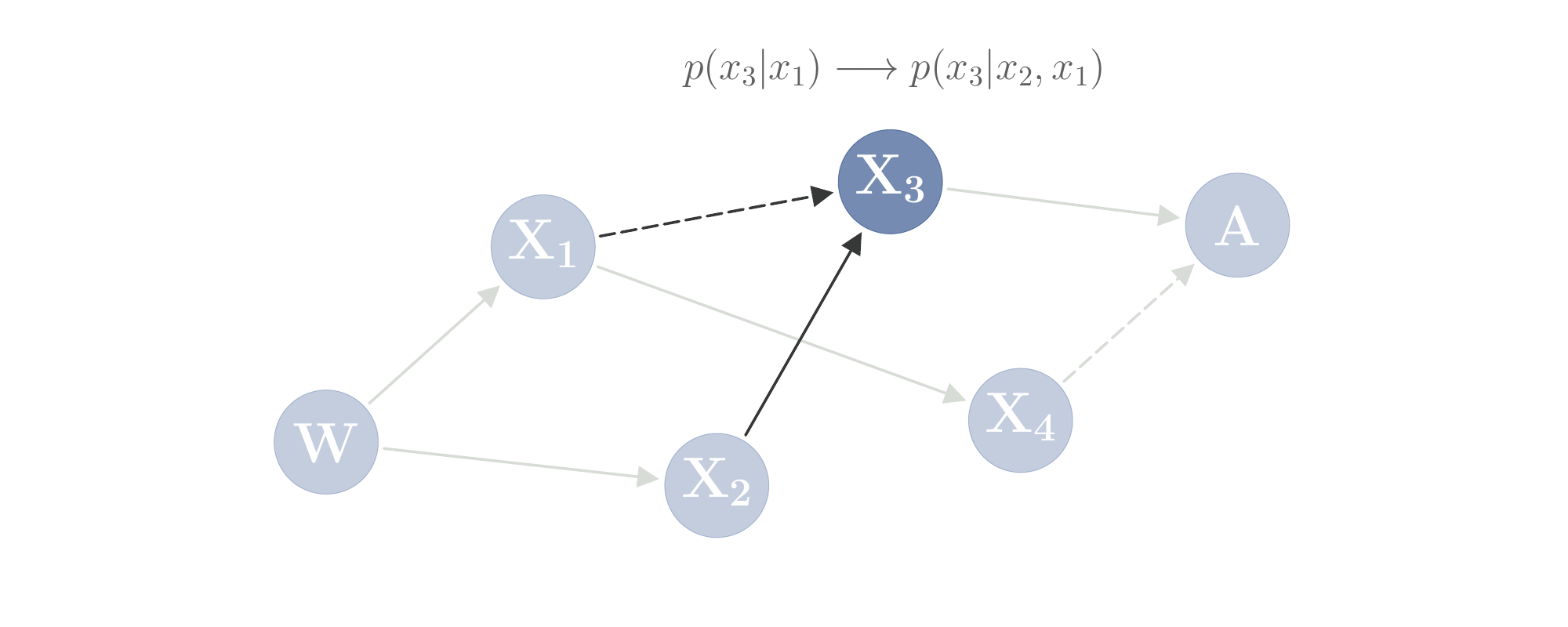}}  
% \begin{center}
% \includegraphics[width = 0.7\textwidth]{example_network.pdf}
% \end{center}
\vspace{-30pt}
\caption{Example of a processing node that is part of a multi-step architecture with $N=5$, visualized as a directed graph. Here, $X_3$ processes the output of $X_2$ by transforming a prior policy $p(x_3|x_1)$ to a posterior policy $P(x_3|x_2,x_1)$. The prior of $X_3$ being conditioned on the output of $X_1$ (indicated by the dashed arrow), means that $X_1$ determines which of the prior policies $\{p(\h\cdot\h |x_1)\}_{x_1\in \mathcal X_1}$ is used by $X_3$ to process a given output of $X_2$.}
\label{fig:example1}
\end{figure}

A specific multi-step architecture is characterized by specifying the explicit dependencies on the preceding variables for each node's prior and posterior, or better the \textit{missing} dependencies. For example, in a given multi-step system, the posterior of the node $X_3$ might depend explicitly on the outputs of $X_1$ and $X_2$ but not on $W$, so that $P(x_3|x_2, x_1,w) = P(x_3|x_2,x_1)$. If its prior has the form $q(x_3|x_1)$, then $X_3$ has to process the output of $X_2$. Moreover, in this case, the actual prior policy $q(\h\cdot\h|x_1)\in\mathbb P_{\mathcal X_3}$ that is used by $X_3$ for decision-making is selected by $X_1$ (see Figure \ref{fig:example1}).

In general, the inputs $X_i,\dots,X_j$ that have to be processed by a particular node $X_k$, are given by the variables in the posterior that are missing from the prior, and, if its prior $q$ is conditioned on the outputs of $X_l,\dots, X_m$, then these nodes select which of the prior policies $\{q(\h\cdot\h|x_l,\dots, x_m)\}_{x_l\in\mathcal X_l,\dots,x_m\in\mathcal X_m} \subset \mathbb P_{\mathcal X_k}$ is used by $X_k$ for decision-making, i.e. for the transformation
\[
q(x_k|x_l,\dots,x_m) \, \longrightarrow \,  P(x_k|x_l,\dots,x_m,x_i,\dots,x_j) \, .
\] 
We denote the collection of \textit{input nodes} of $X_k$ by \smash{$X_{in}^{k}$} ($\coloneqq\{X_i,\dots,X_j\}$) and the \textit{prior selecting nodes} of $X_k$ by \smash{$X_{sel}^{k}$} ($\coloneqq\{X_l,\dots,X_m\}$). The joint distribution of $X_0,\dots,X_N$ is then given by
\begin{equation}\label{joint}
p(x_0,\dots, x_N) = \rho(w) \, P_1\big(x_1\big|x_{sel}^1,x_{in}^1\big) \cdots P_N\big(x_N\big|x^N_{sel},x^N_{in}\big) \, 
\end{equation} 
for all $x_k\in\mathcal X_k$ and $x_{sel}^k\in\mathcal X_{sel}^k$, $x_{in}^k\in \mathcal X_{in}^k$ ($k=1,\dots,N$). 

Specifying the sets $X_{sel}^k$ and $X_{in}^k$ of selectors and inputs for each node in the system then uniquely characterizes a particular multi-step decision-making system. Note that we always have $(X_{sel}^1,X_{in}^1) = (\{\},\{X_0\})$. 

{\color{Black}
Decompositions of the form \eqref{joint} are often visualized by directed acyclic graphs, so-called DAGs \citep[see e.g.][pp.~360]{Bishop2006}. Here, in addition to the decomposition of the joint in terms of posteriors, we have added the information about the prior dependencies in terms of dashed arrows, as shown in Figure \ref{fig:example1}. 
}

\bigskip

\subsection{Multi-step Free Energy principle} \label{sec:nodebasedFE}
If $P_k$ and $q_k$ denote the posterior and prior of the $k$-th node of an $N$-step decision-process, then the Free Energy principle takes the form
\begin{equation}\label{FE:generalq}
\sup_{P_1,q_1,\dots,P_N,q_N}\Big( \langle U\rangle - \sum_{k=1}^N \frac{1}{\beta_k} \, \big\langle\DKL(P_k\|q_k) \big\rangle \Big) \, ,
\end{equation}
where, in addition to the expectation over inputs, the average of $\DKL(P_k\|q_k)$ now also includes the expectation with respect to $X_{sel}$, 
\[
\big\langle \DKL(P\|q)\big\rangle = \sum_{x_{sel},x_{in}} p(x_{sel},x_{in}) \, \DKL\big(P(\h\cdot\h|x_{sel},x_{in})\|q(\h\cdot\h|x_{sel})\big) \, .
\]

Since the prior policies only appear in the KL-divergences, and moreover, there is exactly one KL-divergence per prior, it follows as in \ref{subsec:optprior}, that for each $k=1,\dots,N$ the optimal prior is the marginal given for all $x_k\in\mathcal X_k$ by
\begin{equation} \label{prior:general}
q_k(x_k|x_{sel}) = p_k(x_k|x_{sel}) \coloneqq \frac{1}{p(x_{sel})} \, \sum_{\{x_0,\dots,x_N\}\setminus (\{x_k\}\cup x_{sel}) }  p(x_0,\dots,x_N) \, ,
\end{equation}
whenever $X_{sel}^{k}=x_{sel}$. Hence, the Free Energy principle can be simplified to
\begin{equation}\label{FE:general}
\sup_{P_1,\dots,P_N} \Big( \langle U\rangle - \sum_{k=1}^N \frac{1}{\beta_k} \, I\big(X^{k}_{in};X_k\big|X^{k}_{sel}\big)  \Big) \, ,
\end{equation}
where $I(X;Y|Z)$ denotes the conditional mutual information of two random variables $X,Y$ given a third random variable $Z$.

By optimizing \eqref{FE:general} alternatingly, i.e. optimizing one posterior at a time while keeping the others fixed, we obtain for each $k=1,\dots, N$,
\begin{equation} \label{post:general}
P_k\big(x_k|x_{sel},x_{in}\big) = \frac{p_k\big(x_k|x_{sel}\big)}{Z_k(x_{sel},x_{in})} \, \exp \Big[\beta_k \, \mathcal F_k[P_1,\dots, P_N](x_k,x_{sel},x_{in})\Big] \, ,
\end{equation}
whenever $X_{sel}^{k}=x_{sel}$ and $X_{in}^{k}=x_{in}$. Here, $Z_k(x_{sel},x_{in})$ denotes the normalization constant and $\mathcal F_k[P_1,\dots,P_N]$ denotes the (effective) utility function on which the decision-making in $X_k$ is based on. More precisely, given $\Tilde X = (X_k,X_{sel}^k,X_{in}^k)$, it is the Free Energy of the subsequent nodes in the system, i.e. for any value of $\tilde x\coloneqq (x_k,x_{sel},x_{in})$ we obtain for $\mathcal F_k \coloneqq \mathcal F_k[P_1,\dots, P_N]$,
% Therefore, in general, $\mathcal F_k$ explicitly depends on the full posterior of every node in the system except $P_k$. More precisely, 
\begin{equation} \label{FE:intermediate}
\mathcal F_k(\tilde x) =  \frac{1}{p(\tilde x)} \, \sum_{\{x_0,\dots,x_N\}\setminus \tilde x} p(x_0,\dots,x_N) \, \mathcal F_{k,\mathrm{loc}}(x_0,\dots,x_N) \, ,
\end{equation}
where 
\begin{equation} \nonumber
\mathcal F_{k,\mathrm{loc}}(x_0,\dots,x_N)\coloneqq  U(x_0,x_N) - \sum_{i>k} \frac{1}{\beta_i} \log \frac{P_i(x_i|x^{i}_{sel},x_{in}^{i})}{p_i(x_i|x_{sel}^{i})}  \, .
\end{equation}
Here, $x_{in}^{i}$ and $x_{sel}^{i}$ are collections of values of the random variables in $X_{in}^{i}$ and $X_{sel}^{i}$, respectively. The final Blahut-Arimito-type algorithm consists of iterating \eqref{post:general}, \eqref{prior:general}, and \eqref{FE:intermediate} for each $k=1,\dots,N$ until convergence is achieved. Note that, since each optimization step is convex (\textit{marginal convexity}), convergence is guaranteed but generally not unique \citep{Jain17}, so that, depending on the initialization, one might end up in a local optimum. 

\smallskip

\subsection{Example: two-step information-processing}
The cases of \textit{serial} and \textit{parallel} information-processing studied in \citep{Genewein2013}, are special cases of multi-step decision-making systems introduced above. Both cases are two-step processes ($N=2$) involving the variables $X_0=W$, $X_1=X$, and $X_2=A$. The serial case is characterized by $(X_{sel}^2,X_{in}^2)=(\{\},\{X_1\})$, and the parallel case by $(X_{sel}^2,X_{in}^2)=(\{X_1\},\{X_0\})$. There is a third possible combination for $N=2$, given by $(X_{sel}^2,X_{in}^2) = (\{\},\{X_0,X_1\})$. However, it can be shown that this case is equivalent to the (one-step) rate distoration case from Section \ref{sec:prelim}, because if $A$ has direct world state access, then any extra input to the final node $A=X_2$, that is not a prior selector, contains redundant information.

\section{Systems of bounded rational agents} \label{sec:multi-agent}

\subsection{From multi-step to multi-agent systems} As explained in \ref{sec:singletomulti} above, a single random variable $X_k$ that is part of an $N$-step decision-making system can represent a single agent or a collection of multiple agents, depending on the cardinality of $\mathcal X^k_{sel}$, i.e. whether $X_k$ has multiple priors which are selected by the nodes in $X^k_{sel}$ or not. Therefore, an $N$-step bounded rational decision-making system with $N>1$ represents a \textit{bounded rational multi-agent system} (\textit{of depth $N$}).   

For a given $k\in\{1,\dots,N\}$, each value \smash{$x\in\mathcal X_{sel}^k$} of \smash{$X_{sel}^k$} corresponds to exactly one agent in $X_k$. During decision-making, the agents that belong to the nodes in \smash{$X_{sel}^k$} are choosing which of the \smash{$|\mathcal X_{sel}^k|$} agents in $X_k$ is going to receive a given input $x_{in}$ (see \ref{ex:3step} below for a detailed example). This decision is based on how well the selected agent $x$ will perform on the input $x_{in}$ by transforming its prior policy $p_k(\h\cdot\h|x)$ into a posterior policy $P_k(\, \cdot\, |x,x_{in})$, subject to the constraint
\begin{equation}\label{DKL:ag}
\langle \DKL(P_k||p_k)\rangle(x) \coloneqq \sum_{x_{in}} p(x_{in}|x) \, \DKL\big(P_k(\h\cdot\h|x,x_{in})\| p_k(\h\cdot\h|x) \big) \, \leqslant \, D_x \, ,
\end{equation}
where $D_x>0$ is a given bound on the agent's information-processing capability. Similarly to multi-step systems, this choice is based on the performance measured by the Free energy of the subsequent agents.

\subsection{Multi-agent Free Energy principle} \label{sec:FE_ag}
In contrast to multi-step decision-making, the information-processing bounds are allowed to be functions of the agents instead of just the nodes, resulting in an extra Lagrange multiplier for each agent in the Free Energy principle \eqref{FE:generalq}. As in \eqref{FE:general}, optimizing over the priors yields the simplified Free Energy principle 
\begin{equation}\label{FE:ag}
\sup_{P_1,\dots,P_N} \Big( \langle U\rangle - \sum_{k=1}^N \sum_{x_{sel}^k\in\mathcal X_{sel}^k} \frac{p(x_{sel}^k)}{\beta_k(x_{sel}^k)} \, I\big(X^{k}_{in};X_k\big|X^{k}_{sel}=x_{sel}^k\big)  \Big) \, ,
\end{equation}
which can be solved iteratively as explained in the previous section, the only difference being that the Lagrange parameters $\beta_k$ now depend on $x_{sel}^k$. Hence, for the posterior of an agent that belongs to node $k$, we have
\begin{equation} \label{post:ag}
P_k\big(x_k|x_{sel},x_{in}\big) = \frac{p_k\big(x_k|x_{sel}\big)}{Z_k(x_{sel},x_{in})} \, \exp \Big[\beta_k(x_{sel}^k) \ \mathcal F_k(x_k,x_{sel},x_{in})\Big] \, , 
\end{equation}
where $\beta_k(x_{sel}^k)$ is chosen such that the constraint \eqref{DKL:ag} is fulfilled for all $x\in x_{sel}^k$, and $\mathcal F_k$ is given by \eqref{FE:intermediate} except that now we have 
\begin{equation} \label{FE:loc:ag}
\mathcal F_{k,loc}(x_0,\dots,x_N)\coloneqq  U(x_0,x_N) - \sum_{i>k} \frac{1}{\beta_i(x_{sel}^i)} \log \frac{P_i(x_i|x^{i}_{sel},x_{in}^{i})}{p_i(x_i|x_{sel}^{i})}  \, .
\end{equation}
The resulting Blahut-Arimoto-type algorithm is summarized in Algorithm \ref{alg:ag}.

\begin{algorithm}
\caption{Blahut-Arimito-type algorithm for \eqref{FE:ag}}\label{alg:ag} \label{alg:gen}
\begin{algorithmic}[1]
\Procedure{getMultiagentSolution}{$U,\rho,\{(X_{sel}^k,X_{in}^k)\}_{k=1}^N,\beta,\epsilon$}                      
  \State \textbf{initialize} $p(x_0,\dots,x_N)$ $\forall x_0,\dots,x_N$,  
   \Repeat
   \State $p_0 \gets p$
   \For{$k=1,\dots,N$}
      \settowidth{\maxwidth}{$\mathcal F_{k}(x_k,x_{sel},x_{in})$}
      \State \algalign{$\mathcal F_{k}(x_k,x_{sel},x_{in})$}{$\ \gets \eqref{FE:intermediate}, \eqref{FE:loc:ag}\ \ \forall x_k,x_{sel},x_{in}$} \Comment{\textit{calc.\,effective utility}}
      \State \algalign{$P_k(x_k|x_{sel},x_{in})$}{$\ \gets \eqref{post:ag}\ \ \forall x_k,x_{sel},x_{in}$} \Comment{\textit{update posterior}}
      \State \algalign{$p(x_k|x_{sel})$}{$\ \gets \eqref{prior:general}\ \ \forall x_k,x_{sel}$} \Comment{\textit{update prior}}
      \State \algalign{$p(x_0,\dots,x_N)$}{$\ \gets \eqref{joint}$}{$\ \ \forall x_0,\dots, x_N$} \Comment{\textit{update joint}}

   \EndFor
   \State $error \gets \mathrm{dist}(p,p_0)$
   \Until{$error<\epsilon$}
   \State \textbf{return} $P_1,\dots,P_N$
\EndProcedure
\end{algorithmic}
\end{algorithm}

\subsection{Specialization} \label{sec:spec}
Even though a given multi-agent architecture predetermines the underlying set of choices for each agent, only a small part of such a set might be used by a given agent in the optimized system. For example, all agents in the final step potentially can perform any action $a\in \mathcal A$ (see Figure \ref{fig:ex_3step} and the Example in \ref{ex:3step} below). However, depending on their indiviual information-processing capabilities, the optimization over the agents' priors can result in a (soft) partitioning of the full action space $\mathcal A$ into multiple chunks, where each of these chunks is given by the support of the prior of a given agent $x$, $\mathrm{supp}(p(\h\cdot\h|x))\subset\mathcal A$. Note that the resulting partitioning is not necessarily disjoint, since agents might still be sharing a number of actions, depending on their available information-processing resources. If the processing capability is low compared to the amount of possible actions in the full space, and if there are enough agents at the same level, then this partitioning allows each agent to focus on a smaller number of options to choose from, provided that the coordinating agents have enough resources to decide between the partitions reliably.

Therefore, the amount of prior adaptation of an agent, i.e. by how much its optimal prior $p$ deviates from a uniform prior $p_0$ over all accessible choices, which is measured by the KL-divergence $\DKL(p\|p_0)$, determines its degree of \textit{specialization}. More precisely, we define the specialization of an agent with prior $p$ and choice space $\mathcal X$ by 
\begin{equation}\label{spec}
\mathcal S[p] \coloneqq \frac{\DKL(p\|p_0)}{\log |\mathcal X|}  = 1 - \frac{H[p]}{\log |\mathcal X|}\, ,
\end{equation} 
where $H[p]\coloneqq-\sum_x p(x)\log p(x)$ denotes the Shannon entropy of $p$. By normalizing with $\log|\mathcal X|$, we obtain a quantity between $0$ and $1$, since $0\leqslant H(p)\leqslant \log |\mathcal X|$. Here, $\mathcal S[p]=0$ corresponds to $H[p]=\log|\mathcal X|$, which means that the agent is completely unspecialized, whereas $\mathcal S[p]=1$ corresponds to $H[p] = 0$, which implies that $p$ has support on a single option $x^\ast\in\mathcal X$ meaning that the agent deterministically performs always the same action and therefore is fully specialized.

\begin{figure}[h]
%\nopagenumber
%\renewcommand{\baselinestretch}{1.0}
\hfill
\noindent\makebox[\textwidth]{\includegraphics[width = 1.1\textwidth]{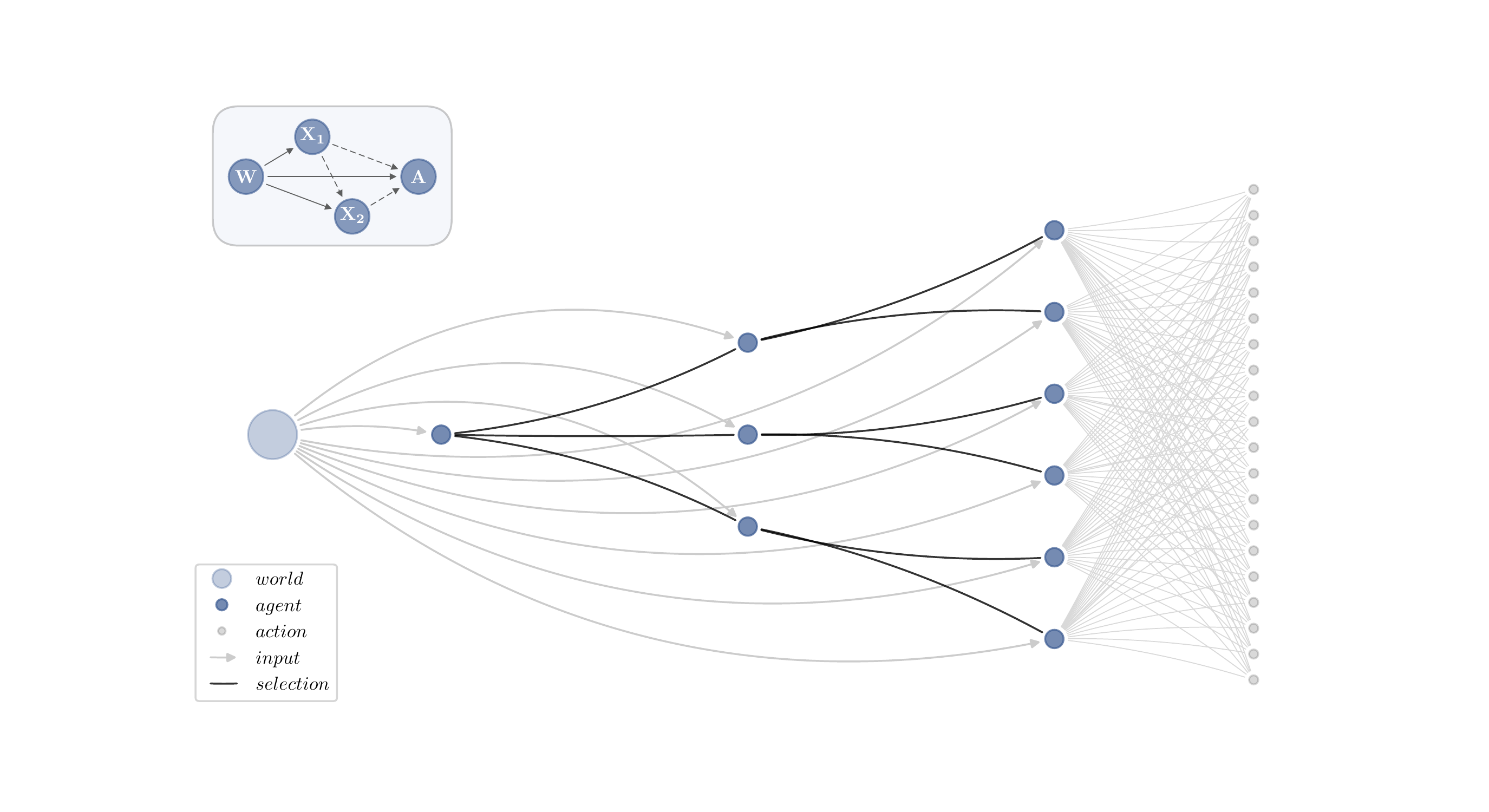}}  
% \begin{center}
% \includegraphics[width = \textwidth]{graph_net_tp14.pdf}
% \end{center}
\vspace{-30pt}
\caption{Example of a hierarchical architecture of 10 agents that are combined via the 3-step decision-making system ($N=3$) shown in the upper left corner (see \ref{ex:3step} for details). Here, every node---and therefore every agent---has access to the world states (big circle). $X_1$ consists of one agent that decides about which of the $|\mathcal X_1|=3$ agents in $X_2$ obtains a given world state as input. The selected agent in $X_2$ selects which of the $|\mathcal X_2|=2$ agents out of the $|\mathcal X_1|\cdot|\mathcal X_2|=6$ agents in $A$ that are connected to it, obtains the world state to perform the final decision about an action $a\in\mathcal A$ (grey circles on the right). In our notation introduced below, this architecture is labelled by $(1,4)_{[1,3,(3,2)]}$ (see Section \ref{sec:characterization}).}
    \label{fig:ex_3step}
\end{figure}

\subsection{Example: Hierarchical multi-agent system with three levels} \label{ex:3step}
Consider the example of an architecture of 10 agents shown in Figure \ref{fig:ex_3step} that are combined via the 3-step decision-making system given by
\begin{equation} \label{ex:3step:def}
(X_{sel}^2,X_{in}^2) = (\{X_1\},\{W\}), \quad (X^3_{sel},X^3_{in}) = (\{X_1,X_2\},\{W\}) , 
\end{equation}
as visualized in the upper left corner of Figure \ref{fig:ex_3step}. The number of agents in each node is given by the cardinality of the target space of the selecting node(s) (or equals \textit{one} if there are no selectors). Hence, $X_1$ consists of one agent, $X_2$ consists of $|\mathcal X_1|$ agents, and $A$ consists of $|\mathcal X_1|\cdot|\mathcal X_2|$ agents. For example, if we have $|\mathcal X_1| = 3$ and $|\mathcal X_2|=2$, as in Figure \ref{fig:ex_3step}, then this results in a hierarchy of $1,3$ and $6$ agents.

The joint probability of the system characterized by \eqref{ex:3step:def} is given by 
\[ 
p(w,x_1,x_2,a) = p(w)P_1(x_1|w)P_2(x_2|x_1,w)P_3(a|x_2,x_1,w) \, ,
\]
and the Free Energy by
\begin{align*}
\mathcal F[P_1,P_2,P_3] \, = & \sum_{w,x_1,x_2,a} p(w,x_1,x_2,a) \bigg[ U(a,w) - \frac{1}{\beta_1} \log\frac{P_1(x_1|w)}{p_1(x_1)} \\
& \quad  - \frac{1}{\beta_2(x_1)}  \log \frac{P_2(x_2|x_1,w)}{p_2(x_2|x_1)} - \frac{1}{\beta_3(x_1,x_2)} \log \frac{P_3(a|x_2,x_1,w)}{p_3(a|x_2,x_1)} \bigg] \, ,
\end{align*}
where the priors $p_1$, $p_2$, and $p_3$ are given by the marginals \eqref{prior:general}, i.e. 
\begin{align*}
p_1(x_1) & \, = \, \sum_{w} \rho(w) P(x_1|w) \, , \\
p_2(x_2|x_1) & \, = \, \sum_w p(w|x_1) P_2(x_2|x_1,w) \, , \\ 
p_3(a|x_2,x_1) & \, = \, \sum_w p(w|x_1,x_2) P_3(a|x_2,x_1,w) .
\end{align*}
By \eqref{post:general}, the posteriors that iteratively solve the Free Energy principle are 
\begin{align*}
P_1(x_1|w)\, & = \, \frac{p_1(x_1)}{Z(w)} \, \exp\big[\beta_1 \mathcal F_1(w,x_1)\big] \, , \\
P_2(x_2|x_1,w)\, & = \, \frac{p_2(x_2|x_1)}{Z(w,x_1)} \, \exp\big[\beta_2(x_1) \mathcal F_2(w,x_1,x_2)\big] \, , \\
P_3(a|x_2,x_1,w)\, & = \, \frac{p_3(a|x_2,x_1)}{Z(w,x_1,x_2)} \, \exp\big[\beta_3(x_1,x_2) U(a,w)\big] \, , 
\end{align*}
where, by \eqref{FE:intermediate} and \eqref{FE:loc:ag},
\begin{align*}
\mathcal F_1(w,x_1) \, & \coloneqq \, \sum_{x_2,a} p(x_2,a|x_1,w)  \Big[U(a,w) - \frac{1}{\beta_2(x_1)} \log \frac{P_2(x_2|x_1,w)}{p_2(x_2|x_1)} \\ 
& \qquad \qquad \qquad \qquad \qquad \ - \frac{1}{\beta_3(x_1,x_2)} \log \frac{P_3(a|x_2,x_1,w)}{p_3(a|x_2,x_1)} \Big] \, ,\\[5pt]
\mathcal F_2(w,x_1,x_2) \, & \, \coloneqq \sum_{a} P_3(a|x_2,x_1,w)  \Big[U(a,w) - \frac{1}{\beta_3(x_1,x_2)} \log \frac{P_3(a|x_2,x_1,w)}{p_3(a|x_2,x_1)}  \Big] \, .
\end{align*}

Given a world state $w\in\mathcal W$, the agent in $X_1$ decides about which of the three agents in $X_2$ obtains $w$ as an input. This narrows down the possible choices for the selected agent in $X_2$ to two out of the six agents in $A$. The selected agent performs the final decision by choosing an action $a\in\mathcal A$. Depending on its degree of specialization, which is a result of his own and the coordinating agents' resources, this agent will choose his action from a certain subset of the full space $\mathcal A$.

\medskip

\section{Optimal Architectures} \label{sec:optimalarchs}

Here, we show how the above framework can be used to determine optimal architectures of bounded rational agents. Summarizing the assumptions made in the derivations, the multi-agent systems that we analyze must fulfill the following requirements: 

\begin{enumerate}[$(i)$]
\item The information-flow is feed-forward: An agent in $X_k$ can obtain information directly from another agent that belongs to $X_m$ only if $m<k$.  
\item Intermediate agents cannot be endpoints of the decision-making process: the information-flow always starts with the processing of $W$ and always ends with a decision $a\in A$.
\item A single agent is not allowed to have multiple prior policies: Agents are the smallest decision-making unit, in the sense that they transform a prior to a posterior policy over a set of actions in one step. 
\end{enumerate}

The performance of the resulting architectures is measured with respect to the expected utility they are able to achieve under a given set of resource constraints. To this end, we need to specify (1) the objective for the full decision-making process, (2) the number $N$ of decision-making steps in the system, (3) the maximal number $n$ of agents to be distributed among the nodes, and (4) the individual resource constraints $\{D_1,\dots,D_n\}$ of those agents. 

{\color{Black}
We illustrate the specifications $(1)$--$(4)$ with a toy example in Section \ref{sec:callcenter} by showcasing and explicitly explaining the differences in performance of several architectures. Moreover, we provide a broad performance comparison in Section \ref{sec:systematic}, where we systematically vary a set of objective functions and resource constraints, in order to determine which architectural features most affect the overall performance. For simplicity, in all simulations we are limiting ourselves to architectures with $N\leqslant 3$ nodes and $n\leqslant 10$ agents. In the following section, we start by describing how we characterize the architectures conforming to the requirements $(i)$--$(iii)$.

\subsection{Characterization of architectures} \label{sec:characterization}
}

\textbf{Type.} In view of property $(ii)$ above, we can label any $N$-step decision-making process by a tuple $(i,j)$, which we call the \textit{type} of the architecture, where $i$ characterizes the relation between the first $N\h{-}\h 1$ variables $W$, $X_1$, \dots, $X_{N-1}$, and $j$ determines how these variables are connected to $X_N=A$.

\begin{figure}
%\nopagenumber
%\renewcommand{\baselinestretch}{1.0}
\hfill
\vspace{-20pt}
\noindent\makebox[\textwidth]{\includegraphics[width = 1.2\textwidth]{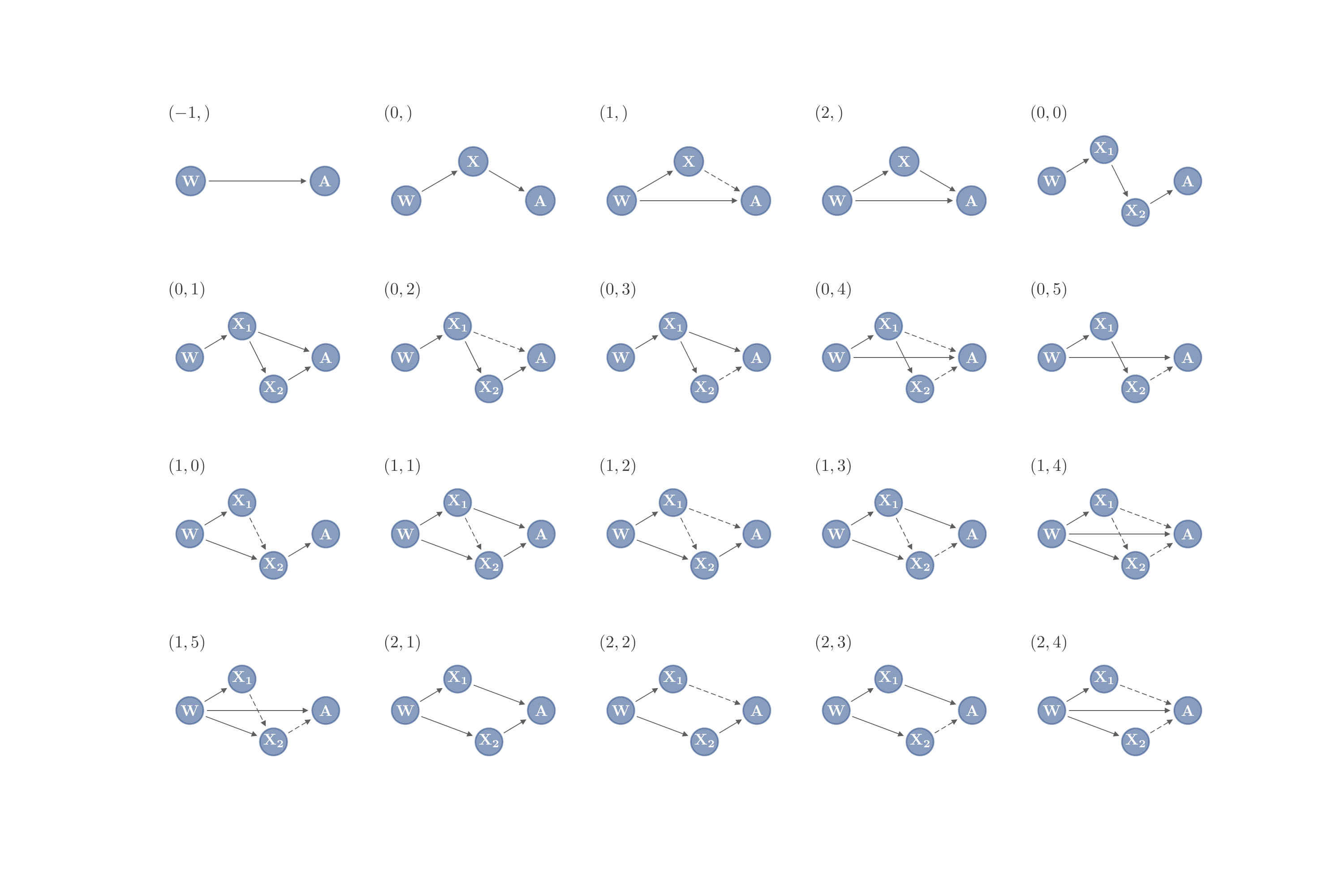}}  
\vspace{-60pt}
\caption{Overview of the resulting architectures for $N\leqslant 3$, each of them being labelled by its type.}
    \label{fig:archs}
\end{figure}

\bigskip 

For example, for $N\leqslant 3$, we obtain the types shown in Figure \ref{fig:archs}, where $i\in\{0,1,2\}$ and $j\in \{0,\dots,5\}$ represent the following relations: 
% The possible network structures crucially depend on the number $N$ of decision nodes in the system. For example, with a maximum of $N\leqslant 3$ steps, we obtain the systems shown in Figure \ref{fig:archs}. We label each of the resulting systems by a \textit{type}, which is given by a tuple $(i,j)$, where $i$ characterizes the relation between the first three variables $W$, $X_1$, and $X_2$, and $j$ characterizes how they are connected to $X_3=A$. More precisely,
\begin{enumerate}[]
\setlength\itemsep{3pt}
\setlength{\itemindent}{20pt}
\item $i=0: \   (X^2_{sel},X^2_{in}) = (\{\},\{X_1\}) $
\item $i=1:  \  (X^2_{sel},X^2_{in}) = (\{X_1\},\{W\}) $
\item $i=2:  \  (X^2_{sel},X^2_{in}) = (\{\},\{W\}) $
\item $j=0:  \  (X^3_{sel},X^3_{in}) = (\{\},\{X_2\}) $
\item $j=1:  \  (X^3_{sel},X^3_{in}) = (\{\},\{X_1,X_2\}) $
\item $j=2:  \  (X^3_{sel},X^3_{in}) = (\{X_1\},\{X_2\}) $
\item $j=3:  \  (X^3_{sel},X^3_{in}) = (\{X_2\},\{X_1\}) $
\item $j=4:  \  (X^3_{sel},X^3_{in}) = (\{X_1,X_2\},\{W\}) $
\item $j=5:  \  (X^3_{sel},X^3_{in}) = (\{X_2\},\{W\})  \, .$
\end{enumerate}
For example, the architecture shown in Figure \ref{fig:ex_3step} has the type $(1,4)$. Correspondingly, the two-step cases are labelled by $(i,)$ for $i\in\{0,1,2\}$, and the one-step rate distoration case by $(-1,)$. Note that not every combination of $i\in\{0,1,2\}$ and $j\in\{0,\dots,5\}$ describes a unique system, e.g. $(2,3)$ is equivalent to $(2,2)$ when replacing $X_1$ by $X_2$. Moreover, as mentioned above, $(2,)$ is equivalent to $(-1,)$, and similarly, $(0,1)$ is equivalent to $(0,)$.

\begin{figure}[h!]
\hfill
\vspace{-40pt}
\noindent\makebox[\textwidth]{\includegraphics[width = 1.2\textwidth]{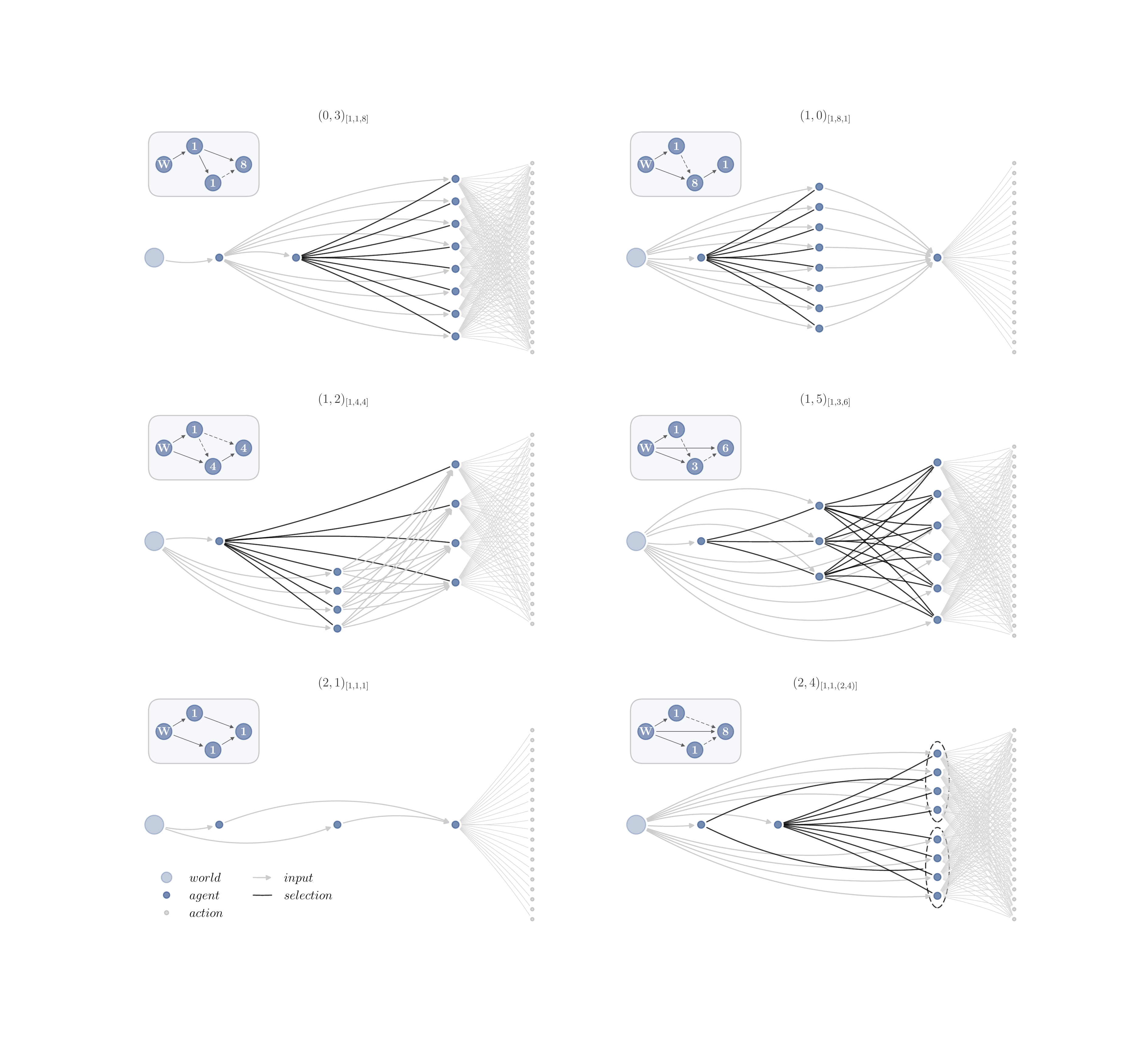}}  
\vspace{-70pt}
\caption{Visualization of exemplary 3-step multi-agent architectures specified by their types and shapes.}
\label{fig:shapes}
\end{figure}

\bigskip

\noindent \textbf{Shape.} After the number of nodes has been fixed, the remaining property that characterizes a given architecture is the number of agents per node. For most architectures there are multiple possibilities to distribute a given amount of agents among the nodes, even when neglecting individual differences in resource constraints. We call such a distribution a \textit{shape}, denoted by $[n_1,n_2,\dots]$, where $n_k$ denotes the number of agents in node $k$. Note that, not all architectures will be able to use the full amount of available agents, most immanently the one-step rate distortion case ($1$ agent), or the two-step serial-case ($2$ agents). For these systems, we always use the agents with the highest available resources in our simulations.

For example, for $N\leqslant 3$ the resulting shapes for a maximum of $n=10$ agents are as follows: 
\begin{itemize}
\setlength\itemsep{0pt}
\item $[1]$ for $(-1,)$, $[1,1]$ for $(0,)$, and $[1,9]$ for $(1,)$, 
\item $[1,1,1]$ for $(0,0)$ and $(2,1)$, 
\item $[1,1,8]$ for $(0,2)$, $(0,3)$, $(0,5)$, $(2,2)$, 
\item $[1,1,(2,4)]$ and $[1,1,(4,2)]$ for $(0,4)$ and $(2,4)$,
\item $[1,8,1]$ for $(1,0)$ and $(1,1)$,
\item $[1,4,4]$ for $(1,2)$,
\item $[1,2,7], [1,3,6], [1,4,5], [1,5,4], [1,6,3], [1,7,2]$ for $(1,3)$ and $(1,5)$,
\item $[1,2,(2,3)]$ and $[1,3,(3,2)]$ for $(1,4)$,
\end{itemize}
where a tuple inside the shape means that two different nodes are deciding about the agents in that spot, e.g. $[1,1,(2,4)]$ means that there are 8 agents in the last node, labeled by the values $(x_1,x_2)\in\mathcal X_1\times\mathcal X_2$ with $|\mathcal X_1| = 2$ and $|\mathcal X_2| = 4$. In Figure \ref{fig:shapes}, we visualize one example architecture for each of the above 3-step shapes, except for the shapes of type $(1,4)$ of which one example is shown in Figure \ref{fig:ex_3step}.

Together, the type $(i,\dots)$ and shape $[n_1,\dots]$ uniquely characterize a given multi-agent \textit{architecture}, denoted by $(i,\dots)_{[n_1,\dots]}$.

\begin{figure}
\hfill
\noindent\makebox[\textwidth]{\includegraphics[width = 0.9\textwidth]{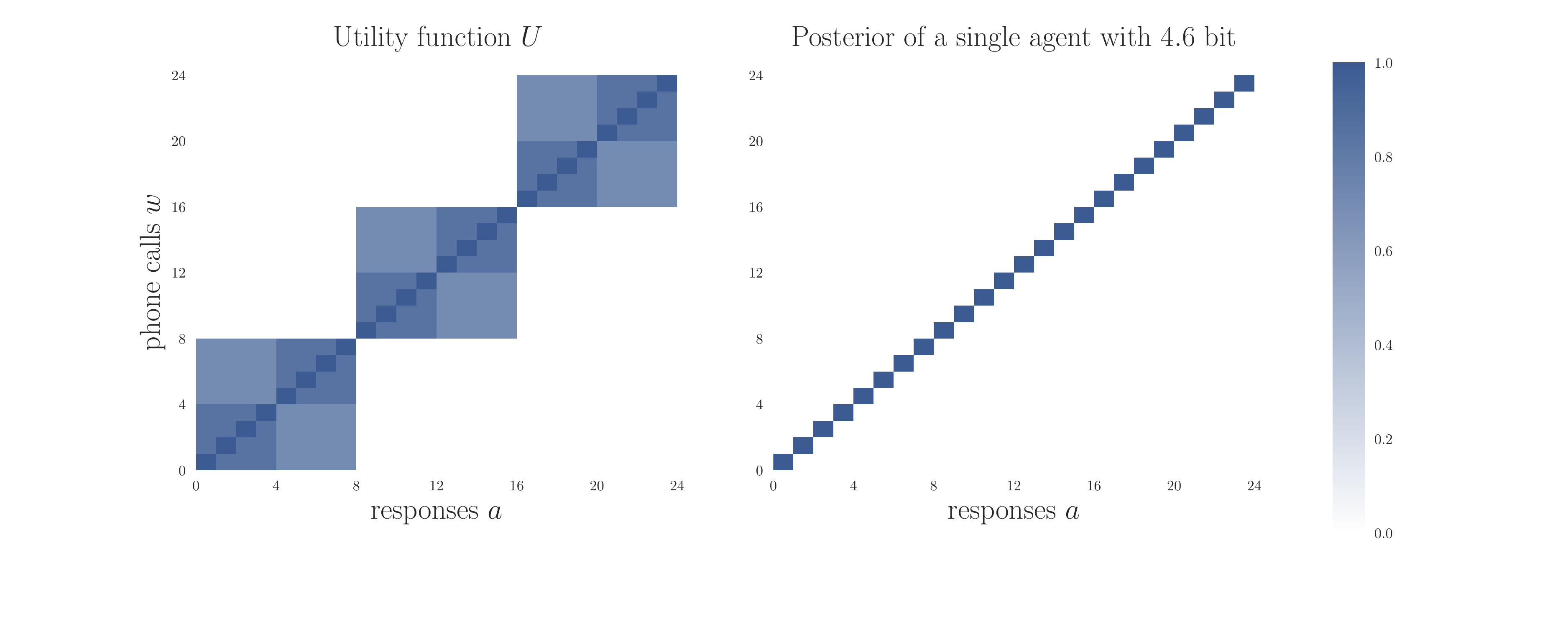}}  
\vspace{-40pt}
\caption{\color{Black}Utility function for Example \ref{sec:callcenter} (left) and posterior policy of a single agent with an information bound of $4.6$ bit (right). The set of phone calls $W$ is partitioned into three separate regions, corresponding to three different topics about which customers might have complaints or questions. Each of these can be divided into two subcategories of four customer calls each. For each phone call there is exactly one answer that achieves the best result ($U=1$). Moreover, the responses that belong to one subcategory of calls are also suitable for the other calls in that particular subcategory, albeit slightly less effective ($U=0.85$) than the optimal answers. Similarly, the responses that belong to the same topic of calls are still a lot better ($U=0.7$) than responses to other topics ($U=0$).}
\label{fig:ex_call_util}
\end{figure}

{\color{Black}
\subsection{Example: Callcenter} \label{sec:callcenter}

Consider the operation of a company's callcenter as a decision-making process, where customer calls (world states) must be answered with an appropriate response (action) in order to achieve high customer satisfaction (utility). The utility function shown in Figure \ref{fig:ex_call_util} on the left can be viewed as a simplistic model for a real-world callcenter of a big company such as a communication service provider. In this simplification, there are 24 possible customer calls that belong to three separate topics, for example questions related to telephone, internet, or television, which can be further subdivided into two subcategories, for example consisting of questions concerning the contract or problems with the hardware. See the description of Figure \ref{fig:ex_call_util} for the explicit utility values.

Handling all possible phone calls perfectly by always choosing the corresponding response with maximum utility requires $\log_2(24) \approx 4.6$ bit (see Figure \ref{fig:ex_call_util}). However, in practice a single agent is usually not capable of knowing the optimal answers to every single type of question. For our example this means that the callcenter only has access to agents with information-processing capability less than $4.6$ bit. It is then required to organize the agents in a way so that each agent only has to deal with a fraction of the customer calls. This is often realized by first passing the phone call through several filters in order to forward it to a specialized agent. Arranging these selector or filter units in a strict hierarchy then corresponds to architectures of the form of $(1,4)$ or $(1,5)$ (see below for a comparison of these two), where at each stage a \emph{single} operator selects how a call is forwarded. In contrast,  architectures of the form of $(2,4)$ allow for multiple independent filters working in parallel, for example realized by multiple trained neural networks, where each is responsible for a particular feature of the call (for example, one node deciding about the language of the call, and another node deciding about the topic). In the following we do not discriminate between human and artificial decision-makers, since both can qualify equally well as information-processing units. 

\begin{figure}[b]
\hfill
\vspace{0pt}
\noindent\makebox[\textwidth]{\includegraphics[width = 1.2\textwidth]{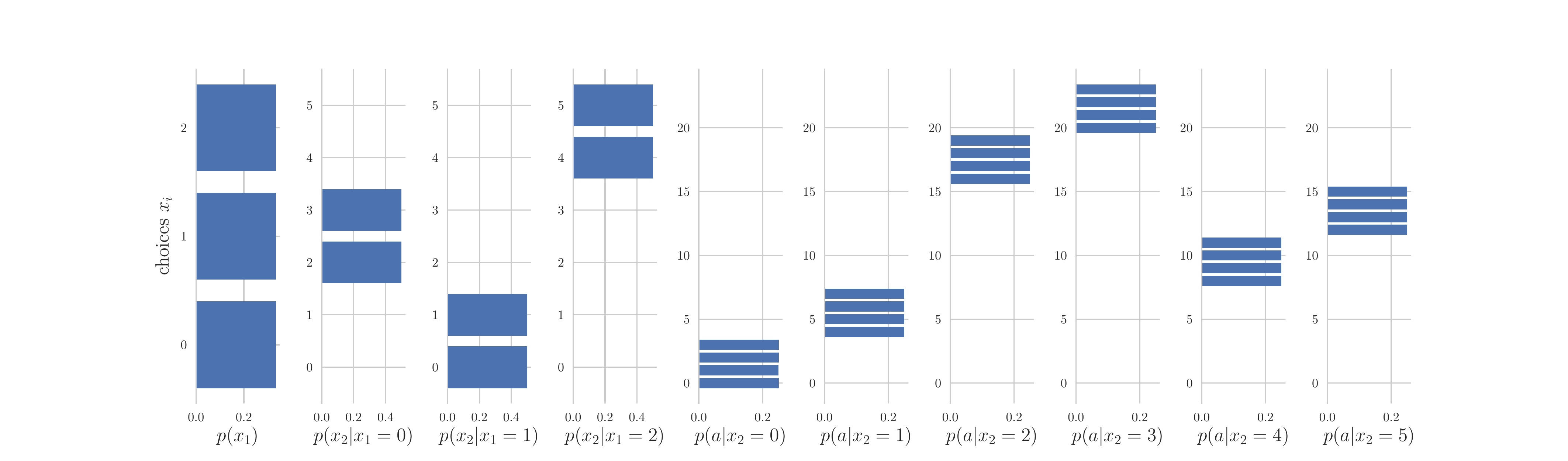}}  
\vspace{-20pt}
\caption{\color{Black}Optimal prior policies for each agent of the architecture $(1,5)_{[1,3,6]}$ with an information bound of $(D_1,D_2,\dots, D_{10}) = (1.6,0.1,\dots,0.1)$.}
\label{fig:ex_call_priors}
\end{figure}

Assume that there are $n=10$ bounded rational agents available. Considering the given utility function, the architectures $(1,4)_{[1,3,(3,2)]}$ (shown in Figure \ref{fig:ex_3step}) and $(1,5)_{[1,3,6]}$ (shown in Figure \ref{fig:shapes}) might be obvious choices as they represent the hierarchical structure of the utility function. With an information bound of $1.6$ ($\approx \log_2(3)$) bit for the first agent and $0.1$ bit for the rest, the optimal prior policies for $(1,5)_{[1,3,6]}$ obtained by our Free Energy principle are shown in Figure \ref{fig:ex_call_priors}. We can see that, for this architecture, the choice $x_1$ of the agent at the first step corresponds to the general topic of the phone call, the decisions $x_2$ of the three agents at the second stage correspond to the subcategory on which one of the six agents at the final stage is specialized to, who then makes the decision about the final response $a$ by picking one of the four actions in the support of its prior.

We can see in Figure \ref{fig:ex_call_comparemany} on the left that a hierarchical structure as in $(1,5)_{[1,3,6]}$ or $(1,4)_{[1,3,(3,2)]}$ is indeed superior when comparing with the architecture $(2,4)_{[1,1,(2,4)]}$, because there is no good selector for the second filter. We have also added two architectures to the comparison that have a bottleneck of the information flow at either end of the decision-making process, $(0,3)_{[1,1,8]}$ and $(1,0)_{[1,8,1]}$ (see Figure \ref{fig:shapes} for a visualization), which are performing considerably worse than the others: in $(0,3)_{[1,1,8]}$ the first agent is the only one who has direct contact to the customer and passes the filtered information on to everybody else, whereas in $(1,0)_{[1,8,1]}$ the customer talks to multiple agents, but these cannot take any decisions but pass on the information to a final decision node who  has to select from all possible options.
Interestingly, as can be seen on the right side of Figure \ref{fig:ex_call_comparemany}, when changing the resource bounds such that the first agent only has $D_1\,{=}\, 1$ bit instead of $1.6$ and the second agent has $D_2\,{=}\,0.5$ bit instead of $0.1$, then the strictly hierarchical architectures $(1,5)_{[1,3,6]}$ and $(1,4)_{[1,3,(3,2)]}$ are outperformed by the architecture $(2,4)_{[1,1,(2,4)]}$, because their first agent is not able to perfectly distinguish between the three topics anymore. This is an ideal situation for $(2,4)_{[1,1,(2,4)]}$, since here the total information-processing for filtering the phone calls is split up efficiently between the first two agents in the system.

\begin{figure}
\hfill
\vspace{-20pt}
\noindent\makebox[\textwidth]{\includegraphics[width = 1.25\textwidth]{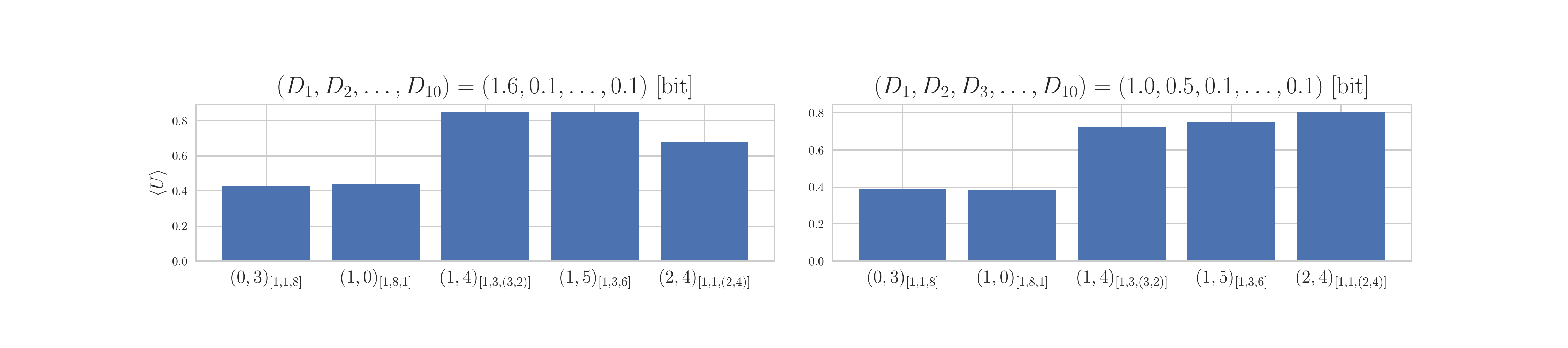}}  
\vspace{-35pt}
\caption{\color{Black}Performance comparison under two different information bounds.}
\label{fig:ex_call_comparemany}
\end{figure}

\begin{figure}[b]
\hfill
\vspace{-0pt}
\noindent\makebox[\textwidth]{\includegraphics[width = 1.22\textwidth]{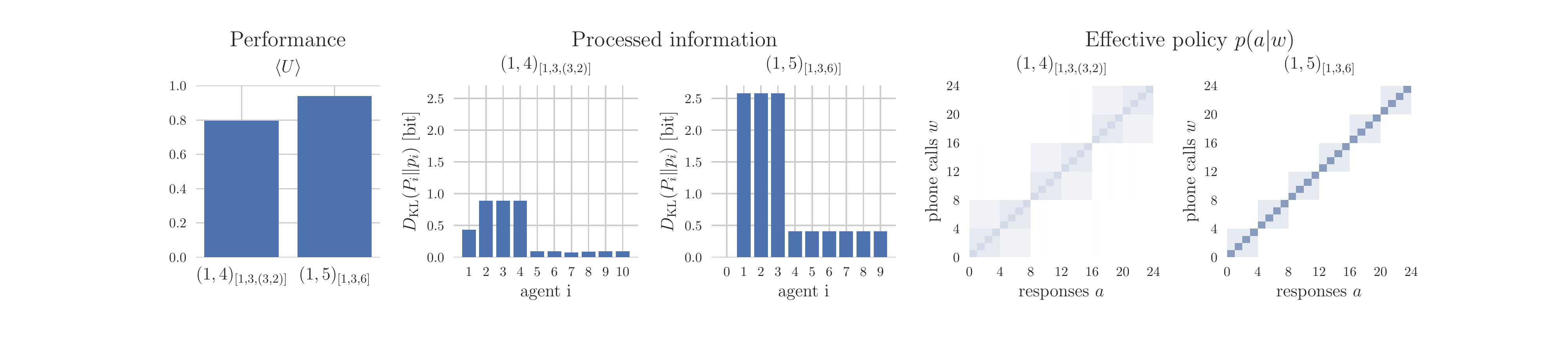}}
\vspace{-30pt}
\caption{\color{Black}Demonstration of the difference between the two architectures $(1,4)_{[1,3,(3,2)]}$ and $(1,5)_{[1,3,6]}$ for an information bound of $D=(0.4,2.6,2.6,2.6,0.4,\dots,0.4)$.}
\label{fig:ex_call_comparepost}
\end{figure}

Note that $(1,4)$ and $(1,5)$ do not necessarily perform identically (as can be seen on the right in Figure \ref{fig:ex_call_comparemany}), even though the structure of the utility function might suggest that it is ideal for $(1,5)_{[1,3,6]}$ to always have the optimal priors shown in Figure \ref{fig:ex_call_priors}. However, this crucially depends on the given information-processing bounds. In Figure \ref{fig:ex_call_comparepost}, we illustrate the difference between the two types in more detail, by showing the processed information that can actually be achieved per agent in the respective architecture for an information bound of $D=(0.4,2.6,2.6,2.6,0.4,\dots,0.4)$. When the first agent in the hierarchy has low capacity, then the rigid structure of $(1,4)$ is penalized because the agents at the second stage cannot compensate the errors of the first agent, irrespectively of their capacity. In contrast, for $(1,5)$, the connection between the second stage and the executing stage can be changed freely, which leads to ignoring the first agent and letting the three agents in the second stage determine the distribution of phone calls completely. In this sense, $(1,5)$ is more robust to errors in the first filter than $(1,4)$.
}

{\color{Black}

\subsection{Systematic performance comparison} \label{sec:systematic}

In this section, we move away from an explicit toy example to a broad performance comparison of all architectures for $N\leqslant 3$, averaged over multiple types of utility functions and a large number of resource constraints (as defined below). In Section \ref{sec:analysis}, this is supplemented with an analysis of the architectural features that best explain the performances.}

\begin{figure}[h]
%\nopagenumber
%\renewcommand{\baselinestretch}{1.0}
\hfill
\vspace{-20pt}
\noindent\makebox[\textwidth]{\includegraphics[width = 1.2\textwidth]{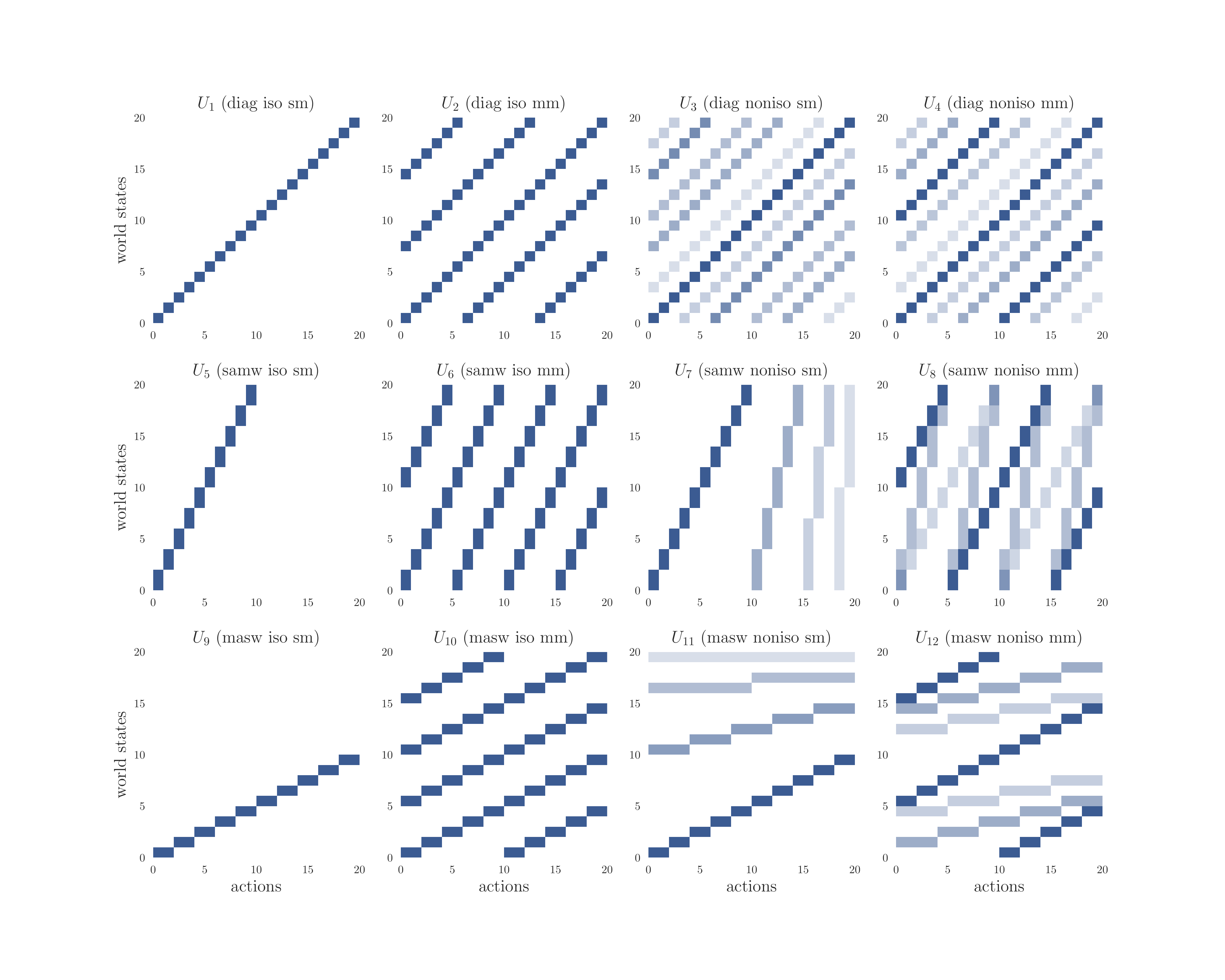}}  
% \begin{center}
% \includegraphics[width = 1.0\textwidth]{utils.pdf}
% \end{center}
\vspace{-50pt}
\caption{Utility functions on which the performances are measured.}
\label{fig:utils}
\end{figure}

\medskip

{\color{Black}
\textbf{Objectives.} \label{sec:objectives} We compare all possible architectures for twelve different utility functions, $\{U_k\}_{k=1}^{12}$, defined on a world and action space of $|\mathcal W| = |\mathcal A| = 20$ elements, and we assume the same cardinality for the range of all hidden variables. Note that the cardinality of the target set $\mathcal X$ for selector nodes $X\in X_{sel}$ is given by the number of agents it decides about. In particular, we consider three kinds of utility functions (one-to-one, many-to-one, one-to-many) that we vary in a $2{\times}2$ paradigm, where the first dimension is the number of maximum utility peaks (single, multiple) and the second dimension is the range of utility values (binary, multi-valued). The utility functions are visualized in Figure \ref{fig:utils}, where the three kinds of functions correspond to the three rows of the plot. A one-to-one scenario applies to a needle-in-a-haystack situation where each world state affords only a unique action, and vice versa each optimal action allows to uniquely identify the world state, for example an absolute identification task. A many-to-one scenario allows for abstractions in the world states, for example in categorization when multiple instances are judged to belong to the same class (e.g. vegetables are boiled, fruit is eaten raw). A one-to-many scenario allows for abstractions in the action space, for example in hierarchical motor control when a grasp action can be performed in many different ways.}

\bigskip

\textbf{Resource limitations.} \label{sec:bounds} We are considering three schemes of resource constraints:
\begin{enumerate}[$(i)$]
\item Same constraints for all agents.
\item Same constraints for all agents but one, which has a higher limit than the other agents.
\item Same constraints for all but two agents, which can have a different limit and have higher limits than all the other agents.
\end{enumerate}
For $(i)$, we compare $20$ sets of constraints $\{D_0,D_1,\dots\}$ with $D_i$ {\color{Black}equally spaced} in the range between $0$ and $3$ bits, for $(ii)$ we compare $39$ sets in the same range but the high resource agent having $1$, $2$ and $3$ bits, and for $(iii)$ we allow $89$ sets with similar constraints than in $(ii)$ but additional combinations for the second high-resource agent.

\begin{figure}
%\nopagenumber
%\renewcommand{\baselinestretch}{1.0}
\hfill
\noindent\makebox[\textwidth]{\includegraphics[width = 1.2\textwidth]{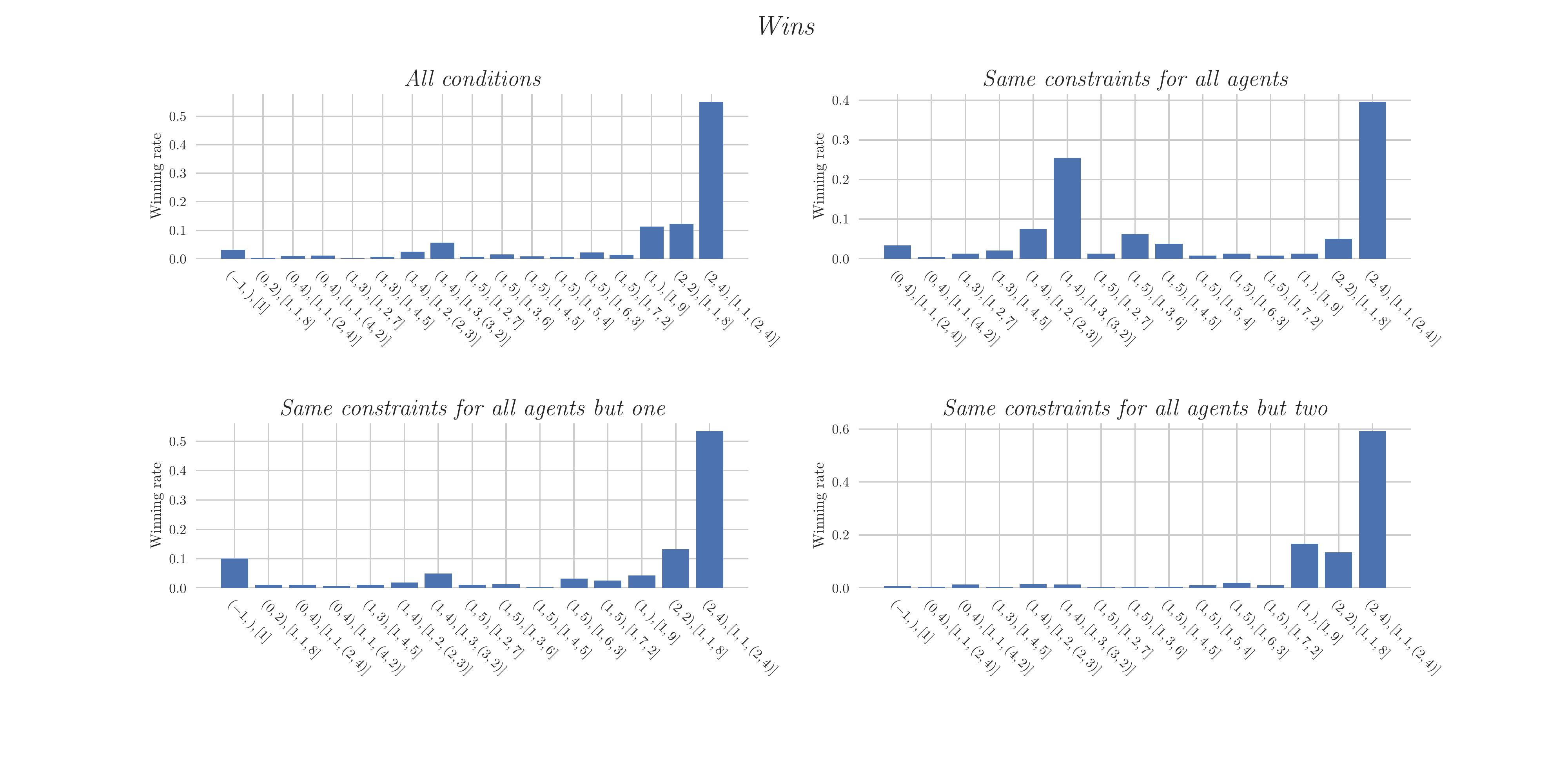}}  
% \begin{center}
% \includegraphics[width = 1.0\textwidth]{wins.pdf}
% \end{center}
\vspace{-50pt}
\caption{Proportion of conditions where the given architectures had the highest performance, for all conditions, and separately for each of the three different schemes of resource constraints.} 
\label{fig:wins} 
\end{figure}

\begin{figure}
%\nopagenumber
% \renewcommand{\baselinestretch}{1.0}
\hfill

\vspace{-110pt}
\noindent\makebox[\textwidth]{\includegraphics[width = 1.1\textwidth]{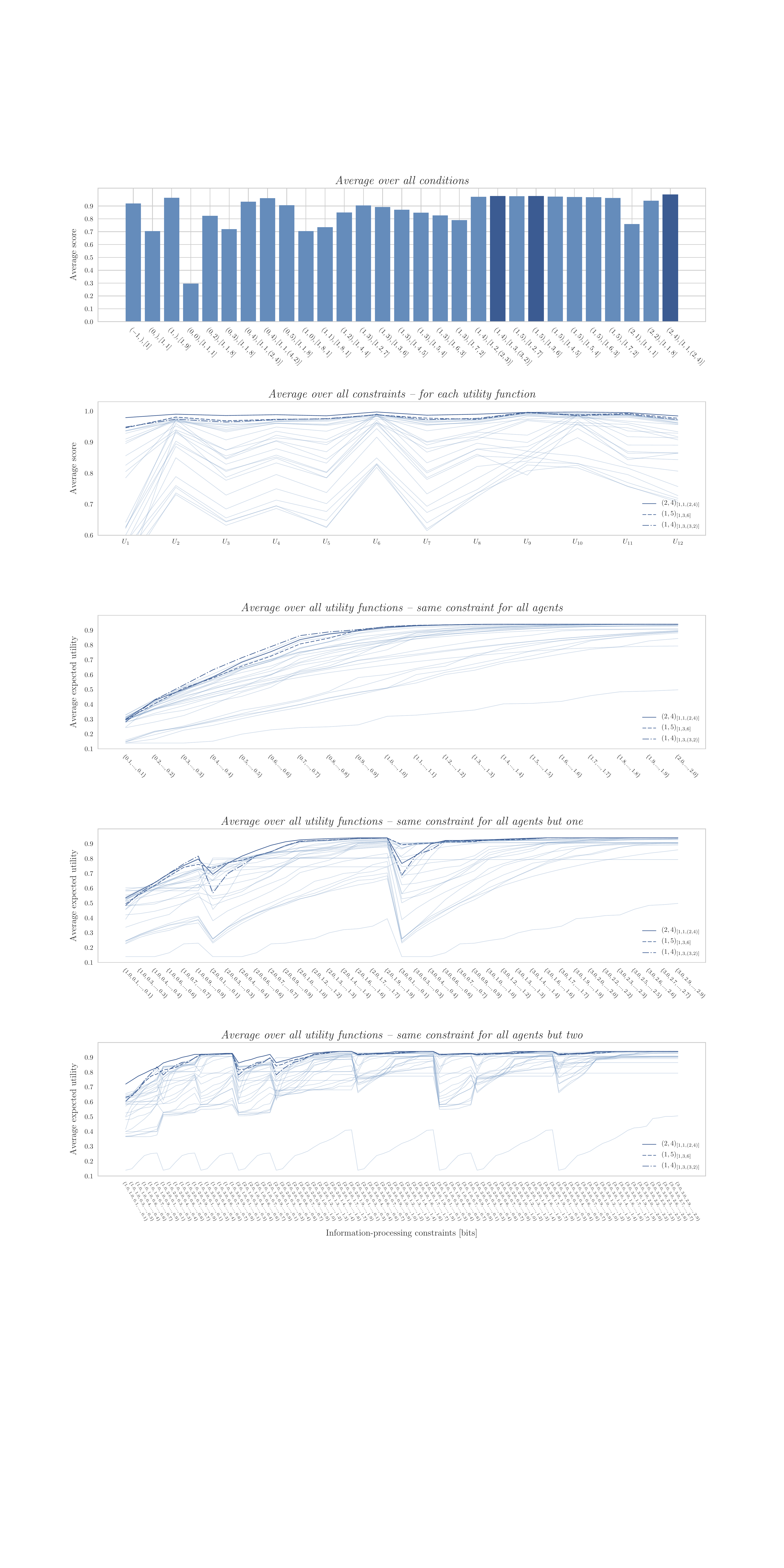}}  
% \begin{center}
% \includegraphics[width = \textwidth]{res_perf.pdf}
% \end{center}
\vspace{-210pt}
\caption{Architecture performances averaged over all conditions (first row), averaged over all information bounds for each utility function (second row), and averaged over all objectives for each information bound (last three rows).} 
\label{fig:results_perf} 
\end{figure}

\bigskip

\textbf{Simulation results.} \label{sec:results} The performance of an architecture is given by its expected utility with respect to a given objective and a given information bound as defined above. In Figure \ref{fig:wins}, we show which of the architectures won at least one condition, together with the proportion of conditions won by each of these architectures. We can see that $(2,4)_{[1,1,(2,4)]}$ overall outperforms all the other systems (see Figure \ref{fig:shapes} for a visualization). In the case when all agents have the same resource constraints, the architecture $(1,4)_{[1,3,(3,2)]}$ is a strong second winner, however this is not the case if one or two agents have more resources than the rest. It is not surprising that in these situations the parallel case with one high-resource agent distributing the work among the low resource agents, and even the case of a single agent that does everything by himself, are both performing well.

% An exemplary result for a specific pair of conditions is shown in Figure \ref{fig:results_ex}. We can see that, in this particular case, the architecture $(1,4)_{[1,3,(3,2)]}$ (see Figure \ref{fig:ex_3step}) performs best. It also shows that the average specialization of the operational agents behaves similarly to the expected utility and thus indicates a possible explanation for the performance of certain architectures. We will come back to this in the next section.  

A closer look on the achieved expected utilities however, shows that there are several architectures that are almost equally well performing for many conditions. In order to increase comparability between the different utility functions, we measure performance in terms of a relative \textit{score}, which, for a given utility function and resource constraint, is given by the architectures' expected utility divided by the maximum expected utility of all architectures. The score averaged over all conditions is shown for each architecture in Figure \ref{fig:results_perf} in the top row. We can see that the best architectures are pretty close to each other. As expected, the architecture that won the most conditions also has the highest overall performance, however there are multiple architectures that are very close. The top three architectures are 
\begin{equation}\label{toparchs}
(2,4)_{[1,1,(2,4)]}, \ (1,5)_{[1,3,6]}, \ (1,4)_{[1,3,(3,2)]}, 
\end{equation}
which have been visualized above (Figure \ref{fig:ex_3step} and \ref{fig:shapes}). 

A better understanding of their performances under different resource constraints can be gathered from the remaining graphs in Figure \ref{fig:results_perf}. In the second row we can see that the top three overall architectures also perform best for almost all utility functions when averaged over the information bounds. The last three graphs in Figure \ref{fig:results_perf} show the expected utility of each architecture averaged over all utility functions for each information bound. We can see how the expected utility increases with higher information bounds, for some architectures more than for others. The top three architecures perform differently for most of the bounds, with spans of bounds where each of them clearly outperforms the others.

\section{Discussion}

\subsection{Analysis of the simulations} \label{sec:analysis}

There are plenty of factors that influence the performance of each of the given architectures. Here, we attempt to unfold the features that determine their performances in the {\color{Black}clearest way}. To this end, we compare the architectures with respect to the following quantities:

\begin{enumerate}[$\quad$]
\setlength\itemsep{3pt}
\item \textit{Average specialization of operational agents}: the specialization \eqref{spec} averaged over all agents in the final stage of the architecture.
\item \textit{Hierarchical}: boolean value that specifies whether an architecture is hierarchical or not, meaning that consecutive nodes are occupied by an increasing amount of agents.
\item \textit{Agents with direct $w$-access}: the number of agents with direct world state access.
\item \textit{operational agents with direct $w$-access}: the number of agents in the last node of the architecture.
\item \textit{Number of $w$-bottlenecks}: the total number of nodes that are missing direct access to the world state. 
\end{enumerate}

\begin{figure}[h!]
%\nopagenumber
%\renewcommand{\baselinestretch}{1.0}
\hfill
\vspace{-60pt}
\noindent\makebox[\textwidth]{\includegraphics[width = 1.2\textwidth]{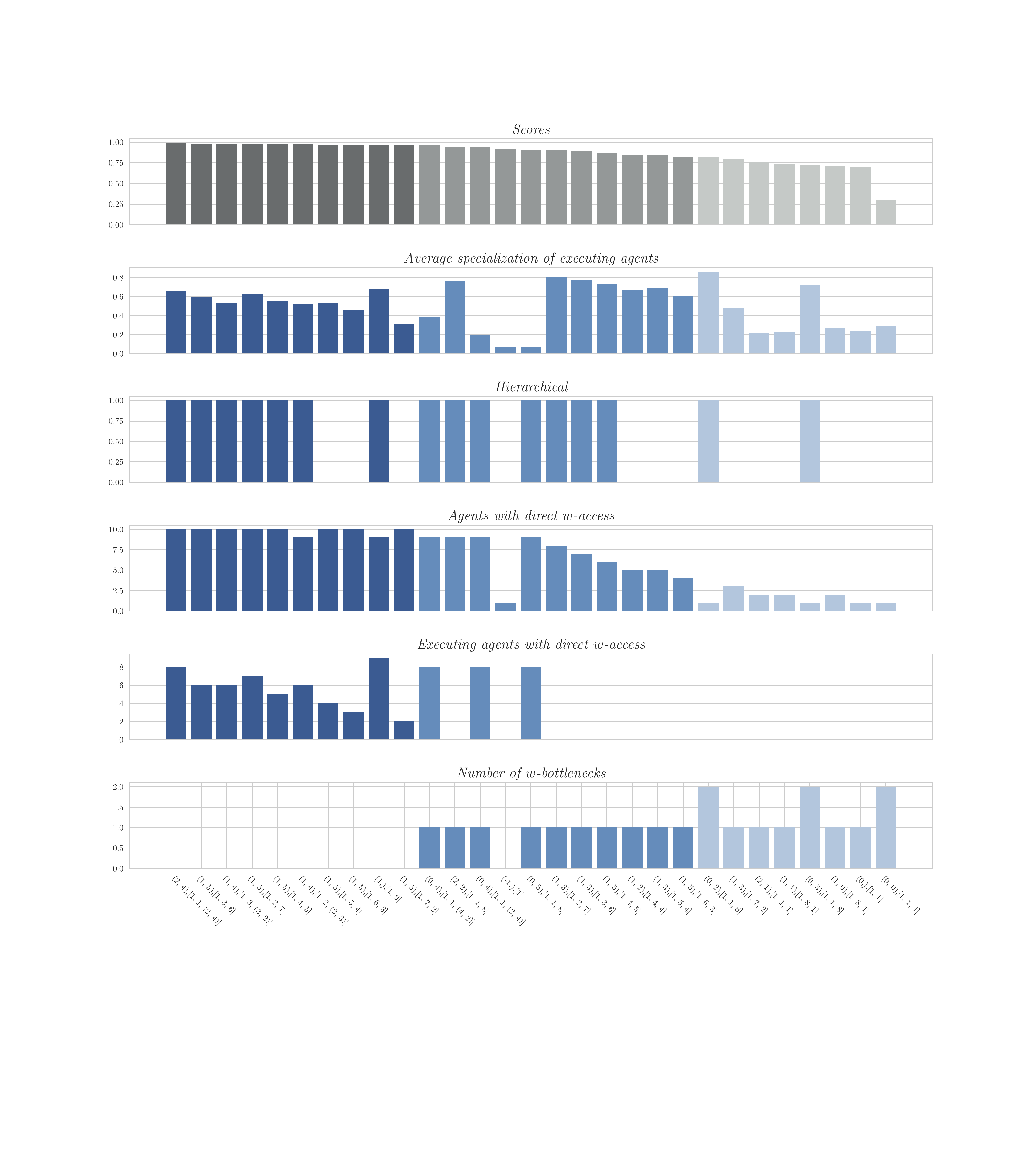}}  
% \begin{center}
% \includegraphics[width = \textwidth]{analysis.pdf}
% \end{center}
\vspace{-120pt}
\caption{Proposed features to explain the architectures' performances (see \ref{sec:analysis}).} 
\label{fig:analysis}
\end{figure}

As can be seen from Figure \ref{fig:analysis}, we found that these {\color{Black}architectural features} explain the differences in performance quite well. More precisely, the architectures can be roughly grouped into three different categories, indicated by slightly different color saturations in Figure \ref{fig:analysis}): The poorest performing group consists of architectures that have between one and two $w$-bottlenecks, and therefore have only few agents with direct $w$-access, in particular none of their operational agents has direct $w$-access. Moreover, in this group, most architectures are not hierarchical at all, and their operational agents have low specialization, with two exceptions that both have two $w$-bottlenecks. 

The architectures with medium performance have maximally one $w$-bottleneck and many of them are hierarchical. Here, those systems that have operational units with high specialization are missing direct $w$-access, and the systems that have operational units with direct $w$-access have low specialization.

All architectures in the top group have many agents with direct world-state access and they have no $w$-bottlenecks. Interestingly, the best six architectures are all strictly hierarchical. Moreover, the order of performance is almost in direct accordance with the average specialization of the operational agents. 

Overall we can say that, it is best to have as many operational units as needed to discriminate the actions well, as long as the coordinating agents have enough resources to discriminate between them properly. The architecture $(1,4)_{[1,1,(2,4)]}$ has eight operational agents, which are managed by two coordinating units, which need maximally two bits (for choosing among four agents) and one bit (for choosing among two agents) in order to perform well. Both of the other top three architectures, $(1,5)_{[1,3,6]}$ and $(1,4)_{[1,3,(3,2)]}$, have six operational agents, which are managed by three coordinating units, so that each of them needs maximally one bit. But compared to $(1,4)_{[1,1,(2,4)]}$, there are less agents to spare for the operational stage. Hence, if the operational units have low resources, it is always a trade-off between the number of operational units and the resources of the coordinating ones.

Another way to see why the architecture $(1,4)_{[1,1,(2,4)]}$ overall outperforms all the other high-ranked systems, might be its lower average \textit{choice-per-agent ratio}, i.e. the average number options for the decision of each agent in the system. In $(1,4)_{[1,1,(2,4)]}$, the second agent also directly observes the world state, and moreover, the choice space of eight agents at the operational stage is split into two and four choices. Therefore, there are only \smash{$\frac{2+4+20}{10} = 2.6 $} choices per agent on average, whereas for \smash{$(1,5)_{[1,3,6]}$} and \smash{$(1,4)_{[1,3,(3,2)]}$}, there are \smash{$\frac{3+6+20}{10} = 2.9$}.

\subsection{Limitations of our analysis}

The analysis presented above only provides a rough explanation of the differences in performance. Which  architecture is optimal, depends a lot on the actual information bounds of each agent. In all of our conditions, we assumed that most agents have the same processing capabilities, which is why there is a certain bias towards architectures that perform well under this assumption (low variance in choice-per-agent ratio across the agents).

Due to the large amount of Lagrange parameters in the Free Energy principle \eqref{FE:ag}, the data generation was done by running the Blahut-Arimoto-type algorithm for $10.000$ different combinations of parameters for each of the architectures, for each type of the different types of resource limitations, $(i)$--$(iii)$ in \ref{sec:bounds}, and for each of the utility functions defined in \ref{sec:objectives}. For a given information bound, the corresponding parameters were determined by looking for the points with the highest Free Energy that still respect the bound.A better approach would be to enhance the global parameter search by a more fine-grained local search. Another possibility is to use an evolutionary algorithm, where each population is given by multiple sets of parameters and the information constraints are built in by a method similar to \citep{Chehouri2016}. This works well but requires significantly more time to process.

Since the Blahut-Arimoto type algorithm is not guaranteed to converge to a global maximum, the resulting values for the expected utility and mutual information for a given set of parameters can depend on the initialization of the algorithm. In practice, this variation is small enough, so that it influences the average performance over multiple conditions only by a negligable amount. However, direct comparisons of architectures for a given information bound and utility function should be repeated multiple times to make sure that the results are stable. 

\bigskip

\subsection{Relation to Variational Bayes and Active Inference} \label{bayesmodelsel}

Above, we determined the architectures that achieve the highest expected utility under a given resource constraint. These constraints are fulfilled by tuning the Lagrange multipliers in the Free Energy principle. If the Lagrange multipliers themselves are fixed, for instance as exchange rates between information and utility \citep{Ortega2010}, or inverse temperatures in thermodynamics \citep{Ortega2013}, then the Free Energy itself would be an appropriate performance measure. {\color{Black}This is done, for example in Bayesian model selection, which is also known as structure learning and represents an important problem in Bayesian inference and machine learning.}
The Bayesian approach for evaluating different Bayesian network structures, in order to find the relation of a given set of hidden variables that best explains a dataset $\mathcal D$, consists in comparing the \textit{marginal likelihood} or \textit{evidence} $p(\mathcal D|S)$ of the structures $S$ \citep{Friedman2003}. This can be seen to be analogous to a performance comparison of different decision-making architectures measured by the Free Energy. In the simple case of one observable $Y$ and one hidden variable $X$, we have
\[
p(y|S) = \sum_{x\in\mathcal X} p(x|S) \, p(y|x,S) \qquad \forall y\in\mathcal Y \, ,
\]
where the likelihood $p(y|x,S)$ is assumed to be known. Given a prior $p(x|S)$ and, for simplicity, a single observed datapoint $y\in Y$, the posterior distribution of $X$ can be inferred by using Bayes' rule,
\begin{equation} \label{bayes}
p(x|y,S) = \frac{p(x|S) \, p(y|x,S)}{p(y|S)}  \qquad \forall x\in\mathcal X \, . 
\end{equation}
{\color{Black}
As has been noted before \citep{Ortega2013}, when comparing \eqref{bayes} with the Boltzmann equation \eqref{boltzmann} we can see that \eqref{bayes} is equivalent to the posterior $P$ of a bounded rational decision-maker with choice space $\mathcal X$, prior policy $p(x|S)$, Lagrange parameter $\beta = 1$, and utility function given by $U(x) \coloneqq \log p(y|x ,S)$. Since the marginal likelihood $p(y|S)$ is the normalization constant in \eqref{bayes}, it follows immediately from \eqref{optFE} that $\log p(y|S)$ is the optimal Free Energy $\mathcal F_{var}[P=p(\h\cdot\h|y,S)]$ of this decision-maker, where
\begin{align} \nonumber
\mathcal F_{var}[P] \, & \coloneqq  \, \sum_x P(x) \log p(y|x ,S)  - \sum_x P(x) \log \frac{P(x)}{p(x|S)} \\ \nonumber
& = \, \sum_x P(x) \log\frac{ p(x|y,S)}{P(x)}+\log p(y|S)\,\\
& = \, \sum_x P(x) \log\frac{ p(x,y|S)}{P(x)} \label{bayes:FE} \, .
\end{align}
In Bayesian statistics, $\mathcal F_{var}$ is known as the \textit{variational} Free Energy, and the given decomposition is often referred to in terms of the difference between accuracy (expected log-likelihood) and complexity (KL-divergence between prior and posterior). It is used in the variational characterization of Bayes' rule, i.e. the approximation of the exact Bayesian posterior $p(\cdot|y,S)$ given by \eqref{bayes} in terms of a simpler---for example a parametrized---distribution $q$ by minimizing the KL-divergence between $q$ and $p(\h\cdot\h|y,S)$. Since $\DKL(q\|p(\h\cdot\h|y,S)) = - \mathcal F_{var}[q] + \log p(y,S)$, this is equivalent to the maximization of $\mathcal F_{var}$. 
}

The same is true for multiple hidden variables. For example, let $S$ be the $3$-step architecture of type $(1,4)$ from Section \ref{ex:3step} with $W = Y$ and hidden variables $X_1,X_2,$ and $X_3=A$. Setting $\beta_1=\beta_2=\beta_3=1$ and $U(a,x_1,x_2,y) = \log p(y|a,x_1,x_2,S)$, we obtain
\[
\mathcal F_2(y,x_1,x_2) = \log p(y|x_1,x_2,S) \, , \ \ \mathcal F_1(y,x_1) = \log p(y|x_1,S) \, ,
\]
and
\[
Z(y,x_1,x_2)=p(y|x_1,x_2,S)\, , \ \ Z(y,x_1) = p(y|x_1,S)\, , \ \  Z(y) = p(y|S) \, .
\]
Note that, even though so far we always assumed that the utility function only depends on the world states and actions, the equations in Sections \ref{sec:multistep}, \ref{sec:multi-agent}, and \ref{ex:3step} are also valid in the general case of $U$ depending on all the variables in the system. The total Free Energy for a given $y\in\mathcal Y$ then takes the form 
\begin{align*} 
& \sum_{x_1,x_2} p(x_1,x_2|y,S) \, \Big( \log p(y|x_1,x_2,S) - \log \frac{p(x_2|x_1,y,S)}{p(x_2|x_1,S)} - \log  \frac{p(x_1|y,S)}{p(x_1|S)} \Big) \\
& = \sum_{x_1}p(x_1|y,S)  \Big( \log p(y|x_1,S)- \log \frac{p(x_1|y,S)}{p(x_1|S)} \Big) \, = \, \log p(y|S) \, .
\end{align*}
Hence, also in this case, the logarithm of the marginal likelihood is given by the Free Energy of the corresponding decision-making system. Choosing the multi-step architecture with the highest Free Energy is then analogous to Bayesian model selection with the marginal likelihood {\color{Black}or Bayesian model evidence} as performance measure. 

{\color{Black}
Another interesting interpretation of \eqref{bayes:FE} is that here the hidden variable $X$ can be thought of as an action causing observed outcomes $y$. This is close to the framework of Active Inference \citep{Friston2015, Friston2017}, where actions directy cause transitions of hidden states, which generate outcomes that are observed by the actor. More precisely, there the real-world process generating observable outcomes is distinguished from an internal generative model describing the beliefs about the external generative process (e.g. a Markov decision process). Observations are generated from transitions of hidden states, which depend on the decision-maker's actions. Decision-making is given by the optimization of a variational Free Energy analogous to \eqref{bayes:FE}, where the log-likelihood is given by the generative model, which describes beliefs about the hidden and control states of the generative process. This way utilities are absorbed into a (desired) prior \citep{Braun2015}. 
There are several differences to our approach. First, the structure of the Free Energy principle of bounded rationality originates from the maximization of a given pre-defined external utility function under information constraints, whereas the Free Energy principle of Active Inference aims to minimize surprise or Bayesian model evidence, effectively minimizing the divergence between approximate and true posterior. Second, in Active Inference, utility is transformed into preferences in terms of prior beliefs, while in bounded rationality prior policies over actions can be part of the optimization process, which results in specialization and abstraction. In constrast, Active Inference compounds utilities and priors into a single desired prior which is fixed and does not allow to separately optimize utility and action priors.
}

\bigskip
\section{Conclusion} \label{sec:conclusion}

In this work, we have presented an information-theoretic framework to study systems of decision-making units with limited information-processing capabilities. It is based on an overreaching Free Energy optimization principle, which, on the one hand, allows to compute the optimal performances of explicit architectures, and on the other hand, produces optimal partitions of the involved choice spaces into regions of specialization. In order to combine a given set of bounded rational agents, first the full decision-making process is split into multiple decision steps by introducing intermediate decision variables, and then a given set of agents is distributed among these variables. We have argued that this leads to two types of agents, non-operational units that distribute the work among subordinates, and operational units that are doing the actual work in the sense of choosing a particular action that either serves as an input for another agent in the system, or represents the final decision of the full process. This ``vertical'' specialization is enhanced by optimizing over the agents' prior policies, which leads to an optimal soft partitioning of the underlying choice space of each step in the system, resulting in a ``horizontal'' specialization as well. 

In order to illustrate the proposed framework, we have simulated and analyzed the performances under a number of different resource constraints and tasks for all possible 3-step architectures whose information flow starts by observing a given world state and ends with the selection of a final decision. Even though the relative architecture performances depend crucially on the explict information-processing constraints, the overall best performing architectures tend to be hierarchical systems of non-operational ``manager'' units at higher hierarchical levels and operational ``worker'' units at the lowest level.

Our approach is based on earlier work on information-theoretic bounded rationality \citep{Ortega2011,Ortega2013,Genewein2013,Genewein2015} (see also the references therein). In particular, the $N$-step decision-making systems introduced in Section \ref{sec:multistep} generalize the two-step processes studied in \citep{Genewein2013,Genewein2015}. According to Simon \citep{Simon1979}, there are three different bounded rational procedures that can transform intractable into tractable decision problems: $(i)$ Looking for satisfactory choices instead of optimal ones, $(ii)$ replacing global goals with tangible subgoals, and $(iii)$ dividing the decision-making task among many specialists. From this point of view, the decision-making process of a single agent, given by the one-step case of information-theoretic bounded rationality \citep{Ortega2011,Ortega2013} described in Section \ref{sec:prelim}, corresponds to $(i)$, while the bounded rational multi-step and multi-agent decision-making processes introduced in Section \ref{sec:multistep} and \ref{sec:multi-agent}, can be attributed to $(ii)$ and $(iii)$.

The main advantage of a purely information-theoretic treatment is its universality. To our knowledge this work is the first systematic theory-guided approach to the organization of agents with limited resources in the generality of information theory. In other approaches, more specific methods are used instead, that are tailored to each particular focus of study. In particular, bounded rationality has usually a very specific meaning, often being implemented by simply restricting the cardinality of the choice space. For example, in management theory the well-known results by Graicunas from the 1930s \citep{Graicunas1933} suggest that managers must have a limited span of control in order to be efficient. By counting the number of possible relationships between managers and their subordinates, he concludes that there is an explicit upper bound of five or six subordinates. Of course, there are many cases of successful companies today that disagree with Graicunas' claim, e.g. Apple's CEO has 17 managers that are reporting directly to him. However, current management experts think that the optimal number is somewhere between 5 and 12. The idea of restricting the cardinality of the space of decision-making is also studied for operational agents. For example in \citep{Camacho1988}, Camacho and Persky explore the hierarchical organization of specialized producers with a focus on production. Even though their treatment is more abstract and more general than many preceeding studies, their take on bounded rationality is very explicit and based on the assumption that the number of elementary parts that form a product, as well as the number of possibilities of each part, are larger than a single individual can handle. Similarly, in most game theoretic approaches that are based on automaton theory \citep{Neyman1985,Rubinstein1988,Hernande2016}, the boundedness of an agent's rationality is expressed by a bound on the number of states of the automaton. Most of these non-information theoretic treatments consider cases when there is a hard upper bound on the number of options, but they usually lack a probabilistic description of the behaviour in cases when the number of options is larger than the given bound. 

The work by \cite{Geanakoplos1991} uses ``information'' to describe the limited attention of managers in a firm. But here, the term is used more informally, and not in the classical information-theoretical sense. However, one of their results suggests that ``firms with more prior information about parameters [...] will employ less able managers, or give their managers wider spans of control'' \cite[p.~207]{Geanakoplos1991}. This observation is in line with information-theoretic bounded rationality, since by optimizing over priors in the Free Energy principle, the required processing-information is decreased compared to the case of non-optimal priors, so that less able agents can perform a given task, or similarly, an agent with a higher information bound can have a larger choice space.

In neuroscience, the variational Bayes approach explained in {\color{Black}Section \ref{bayesmodelsel}} has been proposed as a theoretical framework to understand brain function {\color{Black}in terms of Active Inference \citep{Friston2009,Friston2010, Friston20152,Friston2015, Friston20172, Friston2017}}, where perception is modelled as variational Bayesian inference over hidden causes of observations. There, a processing node (usually a neuron) is limited in the sense that it can only linearly combine a set of input signals into a single output signal. Decision-making is modelled by approximating Bayes' rule in terms of these basic operations, and then tuning the weights of the resulting linear transformations in order to optimize the Free Energy \eqref{bayes:FE}. Hence, there, the Free Energy serves as a tool to computationally simplify Bayesian inference on the neuronal level, whereas our Free Energy principle is a tool to computationally trade off expected utility and processing costs, providing an abstract probabilistic description of the best possible choices when the information-processing capability is limited. 

{\color{Black}
In the general setting of approximate Bayesian inference, there are many interesting algorithms and belief update schemes, for example belief propagation in terms of message passing on factor graphs \citep[see e.g.][]{Yedidia2005}. These algorithms make use of the notion of the Markov boundary (minimal Markov blanket) of a node $X$, which consists of the nodes that share a common factor with $X$ (so-called neighbours). Conditioned on its Markov boundary a given random variable is independent of all other variables in the system, which allows to approximate marginal probabilities in terms of local messages between neighbours. These approximations are generally only exact on tree-like factor graphs without loops \citep[Thm.~14.1]{Mezard2009}. This raises the interesting question of whether such algorithms could also be applied to our setting. First, it should be noted that variational Bayesian inference constitutes only a subclass of problems that can be expressed by utility optimization with information constraints. In this subclass, all random variables have to appear either in utility functions, that is they have to be given as  log-likelihoods, or they have to appear in marginal distributions that are kept fixed---see for example the definition of the utility in the inference example above where $U(a,x_1,x_2,y)=\log p(y|a,x_1,x_2,S)$ compared to the utility functions of the form $U(w,a)$ used throughout the paper that leave all intermediate random variables $X_1,\ldots,X_{N-1}$ unspecified. Second, while it may be possible to exploit the notion of Markov blankets by recursively computing free energies between the nodes in a similar fashion to message-passing, there can also be contributions from outside the Markov boundary, for example when the action node has to take an expectation over possible world states that lie outside the Markov boundary. Finally, it may be interesting to study whether message passing algorithms can be extended to deal with our general problem setting and at least to approximately generate the same kind of solutions as Blahut-Arimoto, even though in general we do not have tree-structured graphs.
}

%% OUTLOOK %% 

There are plenty of other possible extensions of the basic framework introduced in this work. 
\cite{Marschak1998} study multi-agent systems in terms of communication cost minimization, while ignoring the actual decision-making process. 
One could combine our model with the information bottleneck method \citep{Tishby1999} and explicitly include communication costs in order to study more general agent architectures, in particular systems with non-directed information flow. Moreover, we have seen in our simulations that specialization of operational agents is an important feature shared among all of the best performing architectures. In the biological literature, specialization is often paired with modularity. For example \cite{Kashtan2005} and \cite{Wagner2007} show that modular networks are an evolutionary consequence of modularly varying goals. Similarly, it would be interesting to study the effects of changing environments on specialization, abstraction, and optimal network architectures of systems of bounded rational agents.

\bigskip

\section*{Acknowledgement}

This study was funded by the European Research Council (ERC-StG-2015-ERC Starting Grant, Project ID: 678082, ``BRISC: Bounded Rationality in Sensorimotor Coordination'').

\bigskip

\section*{Appendix}

\subsection{Proof of \eqref{post:general}}

The Free Energy functional $\mathcal F$ that is optimized in the Free Energy principle \eqref{FE:general} is given by
\begin{align*}
\mathcal F[P_1,\dots,P_N] =  \sum_x p(x) \, F_{0,\mathrm{loc}}(x) \, ,
\end{align*}
where $x\coloneqq (x_0,\dots,x_N)$, and for all $k\in\{0,\dots,n\}$
\begin{align*}
p(x) & = \Big. \rho(x_0) \, P_1\big(x_1\big|x_{sel}^1,x_{in}^1\big) \cdots P_N\big(x_N\big|x^N_{sel},x^N_{in}\big) \, , \\
\mathcal F_{k,\mathrm{loc}}(x) & =   U(x_0,x_N) - \sum_{i>k} \frac{1}{\beta_i} \log \frac{P_i(x_i|x^{i}_{sel},x_{in}^{i})}{p_i(x_i|x_{sel}^{i})} \, .
\end{align*}
% \begin{align*}
% \frac{\delta \mathcal F}{\delta P_k} (x_k,x_{sel}^k,x_{in}^k) = 
% \end{align*}
By writing 
\[
\mathcal F_{0,\mathrm{loc}}(x) = \mathcal F_{k,\mathrm{loc}}(x) \, - \, \frac{1}{\beta_k} \log \frac{P_k(x_k|x_{sel}^k,x_{in}^k)}{p_k(x_k|x_{sel}^k)} \, -  \, \mathcal R_k(x_{<k})  \, ,
\]
where $x_{<k} \coloneqq (x_0,\dots,x_{k-1})$, and 
\[
\mathcal R_k(x_{<k}) \coloneqq \sum_{i<k} \frac{1}{\beta_i} \log \frac{P_i(x_i|x^{i}_{sel},x_{in}^{i})}{p_i(x_i|x_{sel}^{i})} \, ,
\]
we obtain for any $k\in\{0,\dots,n\}$,
\begin{align*}
\mathcal F[P_1,\dots,P_N]&  = \sum_{x^k_{sel},x^k_{in}} p(x^k_{sel},x^k_{in}) \bigg [ \sum_{x_k} P_k(x_k|x_{sel}^k,x_{in}^k)  \sum_{\tilde x^c} p(\tilde x^c|\tilde x) \mathcal F_{k,\mathrm{loc}} (x) \\
 & \quad  \ - \frac{1}{\beta_k}\DKL \big(P_k(\h\cdot\h|x_{sel}^k,x_{in}^k) \big\| p_k(x_k|x_{sel}^k) \big) \bigg] \, - \, \sum_{x_{<k}} p(x_{<k}) \, \mathcal R_k(x_{<k})
\end{align*}
with $\tilde x = (x_k,x_{sel}^k,x_{in}^k)$ and $\tilde x^c \coloneqq (x_0,\dots,x_N)\setminus \tilde x$. In this form, we can see that optimizing for $P_k$ yields the Boltzmann distribution \eqref{post:general} with respect to the effective utility $\mathcal F_k(\tilde x) = \sum_{\tilde x^c} p(\tilde x^c|\tilde x) \mathcal F_{k,\mathrm{loc}} (x)$ as defined in \eqref{FE:intermediate}.

% \bibliography{../literature_neuro.bib}

\end{document}